\documentclass[format=acmsmall, natbib=false, authordraft=false, screen=true]{acmart}

\RequirePackage[abbreviate=true,
 dateabbrev=true,
 natbib=true,
 isbn=false,
 doi=false,
 eprint=false,
 urldate=comp,
 url=false,
 maxbibnames=9,
 maxcitenames=2,
 backref=false,
 backend=biber,
 style=ACM-Reference-Format,
language=american]{biblatex}

\usepackage{multirow}
\usepackage{booktabs}
\usepackage[inline]{enumitem}
\usepackage{adjustbox}
\usepackage{fixltx2e}
\usepackage{lscape} 

\setcopyright{acmlicensed}
\acmJournal{PACMHCI}
\acmYear{2019} \acmVolume{3} \acmNumber{CSCW} \acmArticle{120} \acmMonth{11} \acmPrice{15.00}\acmDOI{10.1145/3359222}

\received{April 2019}
\received[revised]{June 2019}
\received[accepted]{August 2019}

\acmSubmissionID{cscw120}

\begin{document}

%
\title{Quantifying Voter Biases in Online Platforms: An Instrumental Variable Approach}

%

\author{Himel Dev}
\email{hdev3@illinois.edu}
\affiliation{
  \institution{University of Illinois at Urbana-Champaign}
  \country{USA}
}

\author{Karrie Karahalios}
\email{kkarahal@illinois.edu}
\affiliation{
  \institution{University of Illinois at Urbana-Champaign}
  \country{USA}
}

\author{Hari Sundaram}
\email{hs1@illinois.edu}
\affiliation{
  \institution{University of Illinois at Urbana-Champaign}
  \country{USA}
}


%
\begin{abstract}
In content-based online platforms, use of aggregate user feedback (say, the sum of votes) is commonplace as the ``gold standard'' for measuring content quality. Use of vote aggregates, however, is at odds with the existing empirical literature, which suggests that voters are susceptible to different \emph{biases}---reputation (e.g., of the poster), social influence (e.g., votes thus far), and position (e.g., answer position). Our goal is to quantify, in an observational setting, the degree of these biases in online platforms. Specifically, what are the \emph{causal effects} of different impression signals---such as the reputation of the contributing user, aggregate vote thus far, and position of content---on a participant's vote on content? We adopt an instrumental variable (IV) framework to answer this question. We identify a set of candidate instruments, carefully analyze their validity, and then use the valid instruments to reveal the effects of the impression signals on votes. Our empirical study using log data from Stack Exchange websites shows that the bias estimates from our IV approach differ from the bias estimates from the ordinary least squares (OLS) method. In particular, OLS underestimates reputation bias (1.6--2.2\texttt{x} for gold badges) and position bias (up to 1.9\texttt{x} for the initial position) and overestimates social influence bias (1.8--2.3\texttt{x} for initial votes). The implications of our work include: redesigning user interface to avoid voter biases; making changes to platforms' policy to mitigate voter biases; detecting other forms of biases in online platforms.
\end{abstract}

%
%

%

%

%
\maketitle

\section{Introduction}
In many online platforms, users receive up- and down- votes on content from fellow community members. An aggregate of the votes is commonly used as a proxy for content quality in a variety of applications, such as search and recommendation~\cite{gkotsis2014s, shah2010evaluating, agichtein2008finding, jeon2006framework}. The principle of the \emph{wisdom of the crowds} underlies this quantification, where the mean of judgments on content tends to its true value. The principle rests on the assumption that individuals can make \textit{independent} judgments, and that the crowd comprises agents with \textit{heterogeneous} cognitive abilities~\cite{hong2004groups}.

However, in most online platforms, individuals are prone to using cognitive heuristics that influence their voting behavior and prevent independent judgments~\cite{burghardt2017myopia, burghardt2018quantifying}. These heuristics incorporate different impression signals adjacent to the content---such as the reputation of the contributing user~\cite{beuscart2009distribution}, aggregate vote thus far~\cite{hogg2015disentangling}, and position of content~\cite{lerman2014leveraging}---as input to help individuals make quick decisions about the quality of content. Prior literature suggests that the use of impression signals as shortcuts to make voting decisions results in biases~\cite{lerman2014leveraging, hogg2015disentangling}, where the aggregate of votes becomes an unreliable measure for content quality. We designate these biases as \emph{voter biases}, which stem from the use of impression signals by voters.

There is a plethora of research on detecting and quantifying voter biases in online platforms~\cite{salganik2006experimental, lorenz2011social, krumme2012quantifying, muchnik2013social, lerman2014leveraging, krishnan2014methodology, wang2014quantifying, hogg2015disentangling, stoddard2015popularity, van2016aligning, burghardt2017myopia, abeliuk2017taming, glenski2017rating}. Broadly, researchers have adopted one of the following two approaches: 1) conduct experiments to create different voting conditions for studying participants~\cite{salganik2006experimental, lorenz2011social, muchnik2013social, lerman2014leveraging, hogg2015disentangling, abeliuk2017taming, glenski2017rating}; 2) develop statistical models to analyze historical voting data~\cite{krumme2012quantifying, krishnan2014methodology, wang2014quantifying, stoddard2015popularity, van2016aligning, burghardt2017myopia}. Both approaches have limitations. First, it is hard to perform randomized experiments in actual platforms due to feasibility, ethical issue, or cost~\cite{oktay2010causal}. In addition, researchers not employed at a social media platform are at a disadvantage in conducting such experiments on that platform. Second, statistical models on voter biases often lack causal validity: the derived estimates measure only the magnitude of association, rather than the magnitude and direction of causation required for quantifying voter biases. These limitations of prior research motivate the present work.

\textbf{Present Work.} In this paper, we quantify the degree of voter biases in online platforms. We concentrate on three distinct biases that appear in many platforms---namely, \emph{reputation bias}, \emph{social influence bias}, and \emph{position bias}. Reputation bias captures how the reputation of a user who creates content affects the aggregate vote on that content; social influence bias captures how the initial votes affect the subsequent votes on the same content; position bias captures how the position of content affects the aggregate vote. We study these biases in an observational setting, where we estimate the causal effects of their associated impression signals on the observed votes. 

The key idea of our approach is to formulate voter bias quantification as a causal inference problem. Motivated by the successes of the instrumental variable (IV) framework in studying causal phenomena in Economics and other social science disciplines---e.g., how education affects earnings~\cite{card1999causal}, how campaign spending affects senate elections~\cite{gerber1998estimating}, and how income inequality affects corruption~\cite{jong2005comparative}---we adopt the IV framework to solve our bias quantification problem. The IV framework consists of four components: outcome (dependent variable), exposure (independent variable), instrument (a variable that affects the outcome only through the exposure), and control (other covariates of interest). We operationalize these IV components using variables compiled from log data. We use impression signals as exposure, aggregate feedback as the outcome, and estimate the causal effect of exposure on the outcome by identifying proper instrument and control. 

Identifying an instrument is hard~\cite{angrist2008mostly}. A valid instrument must satisfy three conditions as follows. First, the \emph{relevance} condition requires the instrument to be correlated with the exposure. Second, the \emph{exclusion restriction} requires that the instrument does not affect the outcome directly. Third, the \emph{marginal exchangeability} requires that the instrument and the outcome do not share causes. Of these three conditions, only the relevance condition is empirically verifiable; the remaining two conditions need to be justified through argumentation~\cite{hernan2019causal}. Using large-scale log data from Stack Exchange websites, we identify a set of nuanced instrumental variables for quantifying voter biases. We carefully analyze our proposed instruments to reason about their ability to meet the three instrumental conditions and then select a final set of instruments. We use the final instruments to estimate the causal effects of impression signals on the observed votes using two-stage least squares (2SLS) regression. These regression coefficients provide unbiased causal estimates for quantifying voter biases.

This paper makes the following contributions.
\begin{description}
\item[Bias quantification.] We quantify three types of voter biases by estimating the causal effects of impression signals on the aggregate of votes. Prior research has either used randomized experiments or statistical modeling for quantifying voter biases. While the former can help us identify causal effects, randomized trials are not an option for researchers who work outside the social media platform with observational data. Statistical models help us identify correlation, \textit{not} causation. In contrast, we use an instrumental variable framework by first identifying a set of instrumental variables and then carefully analyzing their validity. The significance of our contribution lies in our framework's ability to identify from observational data, causal factors (impression signals) that affect an individual's vote. 

\item[Findings.] We find that prior work on bias estimation with observational data has significantly underestimated the degree to which factors influence an individual's vote. Our empirical results show that OLS underestimates reputation bias (1.6--2.2\texttt{x} for gold badges) and position bias (up to 1.9\texttt{x} for the initial position), and overestimates social influence bias (1.8--2.3\texttt{x} for initial votes). Furthermore, we find that different impression signals vary in their effect: the badge type (gold, silver, bronze) plays a bigger role in influencing the vote than does reputation score. Also, we find the degree to which each impression signal influences vote depends on the community. This result is significant for two reasons: first, the influence of some of these factors is much more ($\sim$100\% more) than previously understood from statistical models on observational data; despite statistical models estimating regression coefficients, prior work used these coefficients to impute causation, an incorrect inference. Second, had platforms attempted to de-bias with results from prior work, they would have significantly underestimated the effects of reputation and answer position.
\end{description}

\textbf{Significance.} Our identification of causal factors that influence votes has a significant bearing on research in voter bias in particular, as well as the broader CSCW community. First, there are practical implications. Impression signals (answer position, user reputation, prior vote) play a significant role in influencing an individual's vote, at times twice as much as previously understood. Furthermore, the effect of these signals varies by community type (with different content and social norms governing discussions). Second, our work has implications on the future interface design of these platforms. For example, these platforms may conceal impression signals prior to the vote, or delay the vote itself to address social influence bias. Future research is needed, however, to understand the effect of these suggestions. Third, our work informs policy. By identifying causal factors, our work offers social media platforms a way to transparently de-bias votes. The de-biasing may be community dependent. Finally, by introducing the instrumental variable approach to the CSCW community, to identify causal factors from observational data, we hope that more researchers will adopt it to study other questions of interest: e.g., gender and racial bias online.

The rest of this paper is organized as follows. We define our problem in Section \ref{sec:problem} and discuss the related work in Section \ref{sec:related}. We describe our data in Section \ref{sec:data}. We then explain how our method works in Section \ref{sec:method}. Section \ref{sec:results} reports the results of our study. We discuss the implications of our research in Section \ref{sec:discussion} and the limitations in Section \ref{sec:limitations}. Finally, we conclude in Section \ref{sec:conclusion}.
\section{Voter Bias}
\label{sec:problem}
\begin{figure}[htb]
\centering
\includegraphics[scale=0.4]{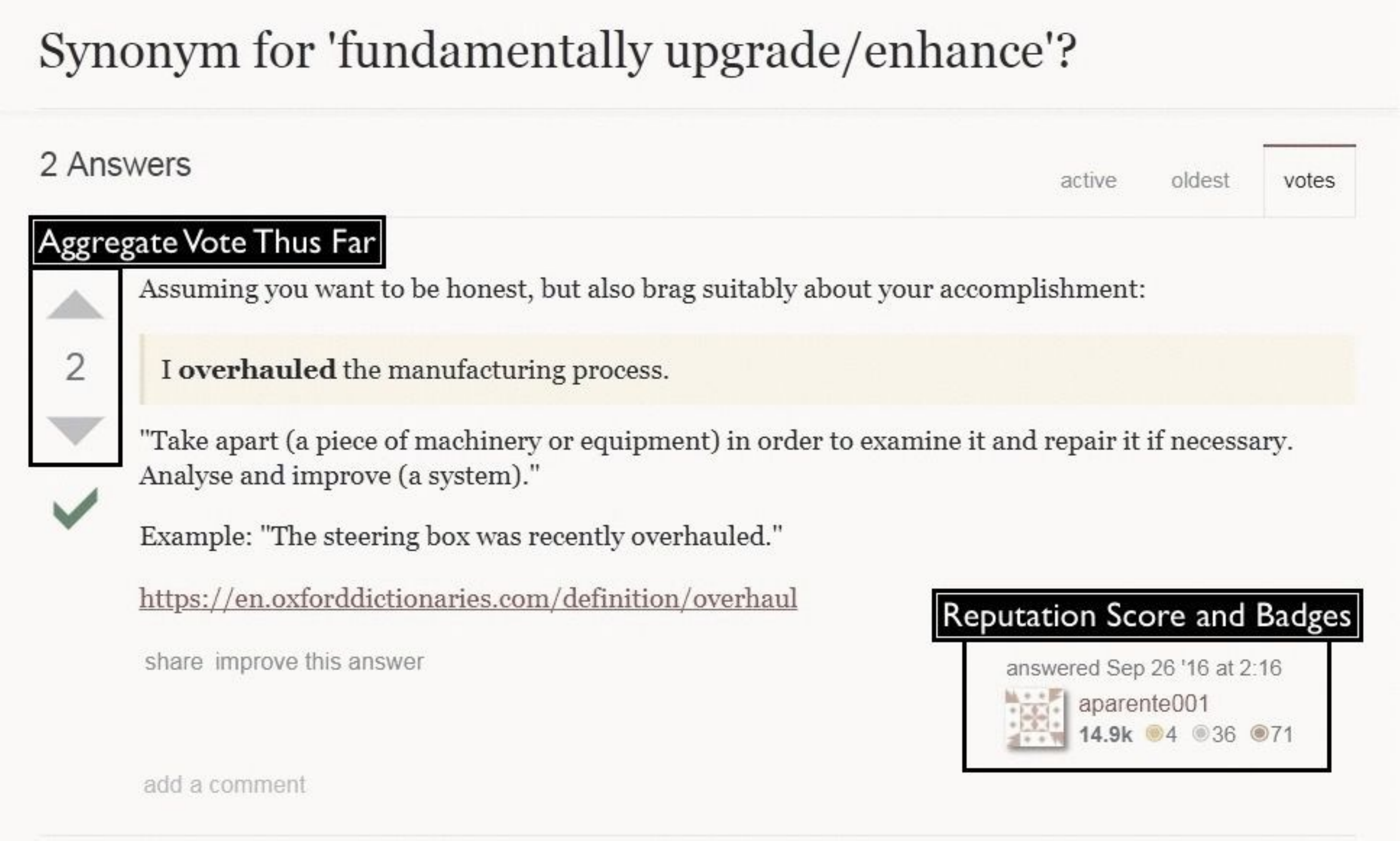}
\caption{A sample page from \textsc{English} Stack Exchange, annotated with different signals that may induce voter biases. For all answers to a question: 1) the score at top left corner shows the aggregate vote thus far (\textit{social signal}), which may induce social influence bias; 2) the statistics at bottom right corner shows the reputation score and badges acquired by the answerer (\textit{reputation signal}), which may induce reputation bias; 3) the answers are presented in a sequential order (\textit{position signal}), which may induce position bias.}
\label{fig:stackexchange_interface}
\vspace{-0.25cm}
\end{figure}

The goal of this paper is to quantify the degree of voter biases in online platforms. We concentrate on three distinct biases: reputation bias, social influence bias, and position bias. To quantify these biases, we estimate the causal effects of their associated signals on the observed votes. In Figure~\ref{fig:stackexchange_interface}, we present a sample page from \textsc{English} Stack Exchange, annotated with different signals that may induce the above-mentioned biases.

\textbf{Reputation Bias.} In content-based platforms (such as Stack Exchange and Reddit), reputation system incorporates the votes on content into the content creator's reputation~\cite{movshovitz2013analysis, richterich2014karma}. In Stack Exchange, for example, votes on content translate into reputation score and badges for the contributing user~\cite{movshovitz2013analysis, anderson2013steering}. The reputation score and badges acquired by each user are visible to all community members, who may use this information to infer the quality of the user's future contributions. Inferring content quality based on user reputation forms the basis of \emph{reputation bias}---when the reputation of a user influences the votes he/she receives on content. We know from prior work that reputation exhibits a Matthew effect~\cite{merton1968matthew}: early reputation increases the chances of future reputation via upvotes. Consider a counterfactual scenario, where two users with different levels of reputation create ``identical'' content; then, reputation bias implies that the user with a higher reputation will receive more upvotes.

\textbf{Social Influence Bias.} The concept of \emph{social influence} in collective action is well-known~\cite{lorenz2011social}: contrary to the \emph{wisdom of the crowds} principle, individuals do not make independent decisions; instead, their decision is influenced by the prior decision of peers. Social influence affects a variety of user activities in online platforms, including voting behavior on content~\cite{stoddard2015popularity, hogg2015disentangling, burghardt2017myopia, burghardt2018quantifying}. Since most platforms reveal the aggregate vote thus far, the initial votes act as a social signal to influence the subsequent voters, forming the basis of \emph{social influence bias}. We know from prior work that for platforms that reveal social signal, users exhibit a herding effect~\cite{glenski2017rating}: the first few votes on content can unduly skew the subsequent votes. Consider a counterfactual scenario, where two ``identical'' content initially receive dissimilar votes; then, social influence bias implies that the content with higher aggregate vote thus far will receive more upvotes.

\textbf{Position Bias.} Many online platforms present content in some order, using a list-style format. For example, in Stack Exchange, answers are sorted based on the aggregate vote thus far. The position of content in a list-style format plays a critical role in deciding how many users will pay attention to it, and interact with it via clicks~\cite{yue2010beyond} or votes~\cite{lerman2014leveraging}. Users pay more attention to items at the top of a list, creating a skewed model of interaction for the items. A consequence of this skewed interaction is \emph{position bias}---when the position of content influences the votes on it. Consider a counterfactual scenario, where two ``identical'' content are located in different positions within a web page; then, position bias implies that the content at the higher position will receive more upvotes.

\textbf{Relationship between Social Influence and Position.} In many platforms, the presentation order of content depends on the aggregate user feedback. In Stack Exchange sites, the default presentation order of answers is the aggregate vote thus far. Quora uses a wide variety of factors to decide the order of answers, including the upvotes and downvotes on the answers. Such vote-dependent ordering scheme imposes a critical challenge in estimating the causal effects of social influence signal and position signal, as the two signals vary together. As such, \emph{the lack of longitudinal variation} in the relationship between the two signals makes it difficult to isolate the effects of their corresponding biases.

\section{Related Work}
\label{sec:related}
Our work draws from and improves upon, a rich literature on online voting behavior and voter biases. Since this paper focuses on quantifying voter biases, we provide a taxonomy of related work on voter biases (Table~\ref{tab:bias_taxonomy}). 

\begin{table}[thb]
\footnotesize\sffamily\mdseries
\caption{A taxonomy of existing literature on voter biases.}
\begin{tabular}{@{}llll@{}}
\toprule
\textbf{Bias} & \textbf{Approach} & \textbf{References} & \textbf{Summary} \\ \midrule

Reputation & Correlation & \cite{beuscart2009distribution}, \cite{pal2011identifying}, \cite{tausczik2011predicting}, & Show some evidence of correlation between past reputation and current \\
Bias & Study &  \cite{paul2012authoritative}, \cite{liang2017knowledge}, \cite{budzinski2018economics} & success~\cite{beuscart2009distribution, pal2011identifying, tausczik2011predicting, paul2012authoritative, liang2017knowledge, budzinski2018economics}. \\ \midrule

Social & Randomized & \cite{salganik2006experimental}, \cite{lorenz2011social}, \cite{muchnik2013social}, & Create different decision making (say voting) conditions for study \\
Influence & Experiment & \cite{abeliuk2017taming}, \cite{glenski2017rating} & participants by varying the availability of preceding decisions~\cite{salganik2006experimental, lorenz2011social, abeliuk2017taming}, \\ 
Bias & & & and purposefully engineered initial decision~\cite{muchnik2013social, glenski2017rating}. \\ \cmidrule{2-4}

& AMT & \cite{hogg2015disentangling}, \cite{celis2016sequential}, \cite{burghardt2018quantifying} & Simulate alternative voting conditions of platform in Amazon Mechanical \\
& Simulation & & Turk (AMT) by varying the availability of preceding decisions~\cite{hogg2015disentangling, celis2016sequential, burghardt2018quantifying}. \\ \cmidrule{2-4}

& Statistical & \cite{krumme2012quantifying}, \cite{krishnan2014methodology}, \cite{wang2014quantifying}, & Develop statistical model for quantifying bias: P\'olya Urn~\cite{krumme2012quantifying, van2016aligning}, \\
& Model & \cite{stoddard2015popularity}, \cite{van2016aligning}, \cite{burghardt2017myopia} &  nonparametric significance test~\cite{krishnan2014methodology}, additive
generative model~\cite{wang2014quantifying}, \\ 
& & &  Poisson regression~\cite{stoddard2015popularity}, logistic regression~\cite{burghardt2017myopia}. \\ \cmidrule{2-4}

& Matching & \cite{wu2008public}, \cite{lederrey2018sheep} & Contrast aggregate user feedback (say ratings) on the same object in \\
& Method & & two different platforms via matching~\cite{lederrey2018sheep}. \\ \midrule

Position & Randomized & \cite{lerman2014leveraging}, \cite{hogg2015disentangling}, \cite{abeliuk2017taming} & Create different decision making (say voting) conditions for study \\
Bias & Experiment & & participants by varying content ordering policies~\cite{lerman2014leveraging, hogg2015disentangling, abeliuk2017taming} \\ \cmidrule{2-4}

& Statistical & \cite{stoddard2015popularity}, \cite{van2016aligning}, \cite{joachims2017unbiased} & Develop statistical model for studying bias: Poisson regression~\cite{stoddard2015popularity}, \\
& Model & & P\'olya Urn~\cite{van2016aligning}, counterfactual inference~\cite{joachims2017unbiased}. \\ \cmidrule{2-4}

& Matching & \cite{radlinski2006minimally}, \cite{yue2010beyond} & Contrast aggregate user feedback (say ratings) for objects occupying \\
& Method & & similar positions~\cite{radlinski2006minimally, yue2010beyond}. \\ 

\bottomrule
\end{tabular}
\label{tab:bias_taxonomy}
\end{table}

\textbf{Voting Behavior.} Recent research has made significant advancements towards the understanding of rating and voting behaviors in online platforms~\cite{gilbert2013widespread, sipos2014review, thebault2017simulation, glenski2017consumers, glenski2018guessthekarma}. Gilbert~\cite{gilbert2013widespread} reported the widespread \emph{underprovisioning of votes on Reddit}: the users overlooked 52\% of the most popular links the first time they were submitted. Using data from Amazon product reviews, Sipos et al.~\cite{sipos2014review} showed that users do not make independent voting decisions. Instead, the decision to vote and the polarity of vote depend on the \emph{context}: a review receives more votes if it is misranked, and the polarity of votes becomes more positive/negative with the degree of misranking. Glenski et al.~\cite{glenski2017consumers} found that most Reddit users do not read the article that they vote on. In a later work, Glenski et al.~\cite{glenski2018guessthekarma} used an Internet game called GuessTheKarma to collect independent preference judgments (free from social and ranking effects) for 400 pairs of images. They found that Reddit scores are not very good predictors of the actual preferences for items as measured by GuessTheKarma. In this paper, we study three distinct cognitive biases that affect user voting behavior. We quantify these biases by estimating the causal effects of their associated signals on the observed votes.

\textbf{Reputation Bias.} Prior works on online reputation suggest that past reputation may be useful in predicting current success~\cite{beuscart2009distribution, pal2011identifying, tausczik2011predicting, paul2012authoritative, liang2017knowledge, budzinski2018economics} (also known as ``superstar economics''~\cite{macdonald1988economics}). Beuscart et al.~\cite{beuscart2009distribution} observed that in MySpace Music, most of the audience is focused on a few stars. These stars are established music artists who signed on major labels. Based on a user study on Twitter, Pal et al.~\cite{pal2011identifying} reported that the popular users get a boost in their authority rating due to the ``name value''. Tausczik et al.~\cite{tausczik2011predicting} found that in MathOverflow, both offline and online reputation are correlated with the perceived quality of contributions. Paul et al.~\cite{paul2012authoritative} found that Quora users judge the reputation of other users based on their past contributions. Liang~\cite{liang2017knowledge} showed that in Reddit, users with higher comment karma tend to produce questions and comments with higher ratings. Budzinski et al.~\cite{budzinski2018economics} analyzed a sample of YouTube stars to show that past success positively and significantly influences current success. While these prior studies show the evidence of reputation bias, they do not provide any bias quantification. In this paper, we provide a quantification of reputation bias through causal estimates.

\textbf{Social Influence Bias.} Since the musiclab experiment by Salganik et al.~\cite{salganik2006experimental}, a large body of work has been devoted to the social influence bias~\cite{wu2008public, lorenz2011social, krumme2012quantifying, muchnik2013social, krishnan2014methodology, stoddard2015popularity, hogg2015disentangling, van2016aligning, abeliuk2017taming, burghardt2017myopia, burghardt2018quantifying}, and its resultant herding effect~\cite{wang2014quantifying, celis2016sequential, glenski2017rating, lederrey2018sheep}. A majority of the work tends to fall into one of two categories---1) Experimental Study: randomized experiment~\cite{salganik2006experimental, lorenz2011social, muchnik2013social, abeliuk2017taming, glenski2017rating}, simulation via Amazon Mechanical Turk (AMT)~\cite{hogg2015disentangling, celis2016sequential, burghardt2018quantifying}; and 2) Observational Study: statistical model~\cite{krumme2012quantifying, krishnan2014methodology, wang2014quantifying, stoddard2015popularity, van2016aligning, burghardt2017myopia}, matching method~\cite{wu2008public, lederrey2018sheep}. Randomized experiments provide a nuanced way to quantify the degree of social influence bias in online platforms; however, often, these experiments are infeasible due to ethical issues, or cost. AMT based simulations fall short in representing the actual voting conditions of a platform. Prior observational studies have used a wide variety of statistical models---P\'olya Urn~\cite{krumme2012quantifying, van2016aligning}, nonparametric significance test~\cite{krishnan2014methodology}, additive generative model~\cite{wang2014quantifying}, Poisson regression~\cite{stoddard2015popularity}, logistic regression~\cite{burghardt2017myopia}---for quantifying social influence bias. However, these studies lack causal validation: the estimates measure only the magnitude of association, rather than the magnitude and direction of causation. For example, in a regression-based herd model, herding behavior could be correlated with the intrinsic quality of content~\cite{van2016aligning}. Therefore, it is difficult to separate the social influence bias from the inherent quality and quantify its effect. In this paper, we adopt the method of instrumental variables to quantify social influence bias.

\textbf{Position Bias.} In recent years, there has been significant interest in studying position bias in online platforms~\cite{radlinski2006minimally, yue2010beyond, lerman2014leveraging, hogg2015disentangling, stoddard2015popularity, van2016aligning, abeliuk2017taming, joachims2017unbiased}. Notably, researchers performed several experimental studies in AMT, where they created different voting conditions for study participants by varying content ordering policies~\cite{lerman2014leveraging, hogg2015disentangling, abeliuk2017taming}. Hogg et al.~\cite{hogg2015disentangling} revealed that social signals affect item popularity about half as much as position and content do. Abeliuk et al.~\cite{abeliuk2017taming} showed that the unpredictability of voting outcome is a consequence of the ordering policy. Lerman et al.~\cite{lerman2014leveraging} found that different policies for ordering content could improve peer recommendation by steering user attention. In this paper, we study position bias in an observation setup, in which it is difficult to isolate the position bias from the social influence bias. To address this problem, we develop a joint IV model that quantifies both position bias and social influence bias.

\section{Data and Variables}
\label{sec:data}
In this section, we first discuss the choice of our data source (Section~\ref{ssec:data_1}), then describe the datasets that we use in this study (Section~\ref{ssec:data_2}); and finally present the variables that we accumulate from the datasets (Section~\ref{ssec:data_3}).

\subsection{Choice of Data Source}
\label{ssec:data_1}
We seek online platforms that satisfy the following criteria: content is user-generated and integral to the platform's success, the position of content and reputation of the contributing user depend upon votes, and the user interface contains various impression signals that may influence the votes. Content-based online platforms such as Quora, Reddit, and Stack Exchange satisfy these criteria. Among them, Reddit and Stack Exchange have publicly available datasets.

We selected the Stack Exchange dataset over Reddit for the following reasons: 1) the Stack Exchange dataset is a complete archive with no missing data (prior work~\cite{gaffney2018caveat} indicates that the Reddit dataset is not complete), which prevents potential selection bias; 2) the governing rules are the same for all Stack Exchange sites (in contrast, subreddits can have different governing rules), which allows us to compare the results across different Stack Exchanges; and 3) the incentives in Stack Exchange sites have been designed for getting to a ``correct'' answer to a question rather than invoking a discussion as is sometimes the case in Reddit, which makes the Stack Exchange content more focused.

\subsection{Stack Exchange Dataset}
\label{ssec:data_2}
Stack Exchange is a network of community question answering websites, where millions of users regularly ask and answer questions on a variety of topics. In addition to asking and answering questions, users can also evaluate answers by voting for them. The votes, in aggregate, reflect the community's feedback about the quality of content and are used by Stack Exchange to recognize the most helpful answers.

\begin{table}[htb]
    \footnotesize\sffamily\mdseries
	\caption{Descriptive statistics for the selected Stack Exchange sites.}
	\begin{tabular}{llrrr}
	\toprule \textbf{Site} & \textbf{Category} & \textbf{\# Users} & \textbf{\# Questions} & \textbf{\# Answers} \\ \midrule
	English & Culture & 169,037 & 87,679 & 210,338\\
	Superuser & Technology & 547,175 & 356,866 & 529,214\\
	Math & Science & 356,699 & 822,059 & 1,160,697\\
	\bottomrule
	\end{tabular}
    \label{tab:data}
\end{table}

\textbf{Use of Published Data.} We obtained Stack Exchange data from \url{https://archive.org/details/stackexchange} on September 2017 (published by Stack Exchange under the CC BY-SA 3.0 license). This snapshot is a complete archive of user-contributed content on the Stack Exchange network. In this paper, we analyze three Stack Exchange sites: \textsc{English}, \textsc{Superuser}, and \textsc{Math}.

\textbf{Inclusion Criteria.} We select the above-mentioned sites for several reasons. First, the three sites represent the three major themes or categories in Stack Exchange: culture [\textsc{English}], technology [\textsc{Superuser}], and science [\textsc{Math}]. Second, apart from \textsc{Superuser}, the remaining two sites are the largest in their category in terms of the number of answers. \textsc{Superuser} is the second largest site in its category, followed by \textsc{StackOverflow}; we discard \textsc{StackOverflow} due to its massive scale difference in comparison to the remaining sites. Third, the sites vary in terms of their susceptibility to voter biases, owing to content that requires interpretation. For example, the quality of answers in \textsc{English} is a lot more subjective compared to the quality of answers in \textsc{Math}. Table~\ref{tab:data} presents descriptive statistics for the three sites analyzed in this paper.

\begin{table}[thb]
\footnotesize\sffamily\mdseries
\caption{The description of variables used in this study. The variables fall into four groups based on the following constructs: site (the Stack Exchange site), question (the question that has been addressed by the answer), answer (the answer in consideration),  and answerer (the user who created the answer).}
\begin{tabular}{@{}llll@{}}
\toprule
\textbf{ID} & \textbf{Variable} & \textbf{Description} \\ \midrule
V\textsubscript{1} & \textnormal{Site} & The Stack Exchange site in consideration \\ \midrule
V\textsubscript{2} & T & The limiting time of bias formation specific to the question \\
V\textsubscript{3} & QuestionViewCount & Number of users who viewed the question \\
V\textsubscript{4} & QuestionFavoriteCount & Number of users who favorited the question \\
V\textsubscript{5} & QuestionScore & Aggregate vote (total upvotes - total downvotes) on the question \\
V\textsubscript{6} & QuestionScoreT- & Aggregate vote on the question before time T \\
V\textsubscript{7} & QuestionScoreT+ & Aggregate vote on the question after time T \\
V\textsubscript{8} & QuestionCommentCount & Number of comments on the question \\
V\textsubscript{9} & QuestionCommentCountT- & Number of comments on the question before time T \\
V\textsubscript{10} & QuestionCommentCountT+ & Number of comments on the question after time T \\
V\textsubscript{11} & QuestionAnswerCount & Number of answers to the question \\
V\textsubscript{12} & QuestionAnswerCountT- & Number of answers to the question before time T \\
V\textsubscript{13} & QuestionAnswerCountT+ & Number of answers to the question after time T \\ \midrule
V\textsubscript{14} & AnswerDayOfWeek & The day of answer creation \\
V\textsubscript{15} & AnswerTimeOfDay & The time of answer creation \\
V\textsubscript{16} & AnswerEpoch & Time gap between between the 1st post in site and the answer \\
V\textsubscript{17} & AnswerTimeliness & Time gap between the question and the answer \\
V\textsubscript{18} & AnswerOrder & Chronological order of the answer \\
V\textsubscript{19} & AnswerScore & Aggregate vote on the answer \\
V\textsubscript{20} & AnswerScoreT- & Aggregate vote on the answer before time T \\
V\textsubscript{21} & AnswerScoreT+ & Aggregate vote on the answer after time T \\
V\textsubscript{22} & AnswerPosition & Position of the answer based on the aggregate vote \\
V\textsubscript{23} & AnswerPositionT- & Position of the answer based on the aggregate vote before time T \\
V\textsubscript{24} & AnswerPositionT+ & Position of the answer based on the aggregate vote after time T \\
V\textsubscript{25} & AnswerCommentCount & Number of comments on the answer \\
V\textsubscript{26} & AnswerCommentCountT- & Number of comments on the answer before time T \\
V\textsubscript{27} & AnswerCommentCountT+ & Number of comments on the answer after time T \\ \midrule
V\textsubscript{28} & AnswererPostCount & Number of posts (questions and answers) written by the answerer \\
V\textsubscript{29} & AnswererAnswerCount & Number of answers written by the answerer \\
V\textsubscript{30} & AnswererActiveAge & Time gap between between the answerer's 1st post and the answer \\
V\textsubscript{31} & AnswererReputation & Total score of questions and answers written by the answerer \\
V\textsubscript{32} & AnswererReputationViaAnswer & Total score of answers written by the answerer \\
V\textsubscript{33} & AnswererGoldCount & Number of gold badges acquired by the answerer \\
V\textsubscript{34} & AnswererSilverCount & Number of silver badges acquired by the answerer \\
V\textsubscript{35} & AnswererBronzeCount & Number of bronze badges acquired by the answerer \\
V\textsubscript{36} & AnswererBadgeDistrribution & {[}AnswererGoldCount, AnswererSilverCount, AnswererBronzeCount{]} \\
V\textsubscript{37} & AnsweredQuestionViewTotal & Total number of users who viewed past questions answered by the answerer \\
V\textsubscript{38} & AnsweredQuestionFavoriteTotal & Total number of users who favorited past questions answered by the answerer \\
V\textsubscript{39} & AnsweredQuestionScoreTotal & Total score of past questions answered by the answerer \\
V\textsubscript{40} & AnsweredQuestionCommentTotal & Total number of comments on past questions answered by the answerer \\
V\textsubscript{41} & AnsweredQuestionAnswerTotal & Total number of answers to past questions answered by the answerer \\ \bottomrule
\end{tabular}
\vspace{-5mm}
\label{tab:var}
\end{table}

\subsection{Variables}
\label{ssec:data_3}
In Stack Exchange sites, questions and answers are the primary content. Answer quality is especially important for these platforms as they thrive to provide answers. For this reason, we analyze the votes on answers. We compile a wide range of variables to capture the voter biases, the factors related to these biases, and the potential effects of these biases. Table~\ref{tab:var} describes the variables used in this study.

\section{Method}
\label{sec:method}
In this section, we first discuss our choice of method for voter bias quantification (Section~\ref{ssec:method_1}), then explain the fundamentals of the chosen method (Section~\ref{ssec:method_2}); and finally present our models for quantifying voter bias (Section~\ref{ssec:method_3} and~\ref{ssec:method_4})

\subsection{Choice of Method}
\label{ssec:method_1}
The goal of this paper is to quantify the degree of voter biases in online platforms. To determine these biases, we need to estimate the \emph{causal effects} of different impression signals on the observed votes. Estimating causal effects from observational data is exceptionally challenging~\cite{winship1999estimation}. The main reason is that there may exist hidden confounders that affect both independent (say impression signal) and dependent (observed votes) variables. A hidden confounder may explain the degree of association between the variables, which prevents standard regressions methods from providing causal estimates~\cite{winship1999estimation, angrist2008mostly}. We observe that our voter bias quantification problem is susceptible to several hidden confounders, such as the quality of the content (from the perspective of voters) and the ability of users (to generate high-quality content). These confounders (e.g., the ability of users) may affect both the impression signals (e.g., the reputation of the contributing user) and the observed votes. Ergo, we need to eliminate the effects of these confounders for estimating the causal effect.

The instrumental variable (IV) approach has been successfully used in the social sciences~\cite{card1999causal, gerber1998estimating, jong2005comparative} to estimate causal effects (e.g., how education affects earning~\cite{card1999causal}, how campaign spending affects senate selection~\cite{gerber1998estimating}, and how income inequality affects corruption~\cite{jong2005comparative}) from observational data. The IV method is especially useful for estimating effects in the presence of hidden confounders~\cite{angrist2008mostly, hernan2019causal}. The technique requires identifying candidate instruments that are correlated with the independent variable of interest. It then relies on careful argumentation (thought experiments) to eliminate the candidate instruments that may affect the hidden confounders. This process implies that the remaining instruments co-vary only with the independent variable, and cannot influence the dependent variable through a hidden confounder. As such, instrumental variables allow us to estimate causal effects, even in the presence of hidden confounders.

Prior research on voter biases regress aggregate vote on impression signals using ordinary least squares (OLS) and interpret the regression coefficients as effects. However, OLS only captures the correlation among variables; the resultant estimates are \emph{non-causal}. For instance, a positive OLS estimate corresponding to an impression signal does not imply that the signal has a positive effect on the aggregate vote; the effect could be zero or even negative. This argument is especially applicable in the presence of hidden confounders. In fact, in such a case, the OLS estimate is biased~\cite{angrist2008mostly}.

\begin{table}[hbt]
\footnotesize\sffamily\mdseries
\caption{The parallels between voter bias quantification and instrumental variable method.}
\begin{tabular}{@{}lll@{}}
\toprule
\textbf{IV Terminology} & \textbf{Bias Terminology} & \textbf{Example} \\ \midrule
Outcome & Aggregate Feedback & Mean of votes on content \\
Exposure & Impression Signal & Reputation of the contributing user \\
Confounder & Unobserved Quality &  What a voter assesses the quality of the content to be \\
Regression Coefficient & Voter Bias & How the reputation of the contributing user affects the mean vote \\ \bottomrule
\end{tabular}
\label{tab:bias_to_iv}
\end{table}

The key conceptual difference between the IV and OLS is: IV relies on argumentation to reason about the underlying causal structure. If all we have access to is observational data, then careful argumentation is necessary to establish the causal structure. As pointed out by Judea Pearl, \emph{``behind every causal conclusion there must lie some causal assumption that is not testable in observational studies''}~\cite{pearl2001bayesianism}. As we can not conduct randomized control trials on the actual platforms, and only have access to the observational data, IV is a reasonable approach for estimating causal effects. Further, our problem aligns well with the use case of IV: estimating causal effect in the presence of hidden confounders (In Table~\ref{tab:bias_to_iv}, we show the parallels between our problem and IV). For these reasons, we adopt the IV method to quantify voter biases.

\subsection{Instrumental Variable Estimation}
\label{ssec:method_2}
To motivate the use of IVs, we now explain a classic well-understood example: the causal effect of education on earnings~\cite{card1999causal}. In general, education enables individuals to earn more money, say through employment that is reserved for college graduates. One can estimate the return to education by simply regressing the earnings of individuals on their education level. However, this simplistic approach has a major limitation in the form of omitted variable---\emph{the unobserved ability of individuals}. Unobserved ability (\emph{confounder}) might be correlated with the level of education that an individual attains (\emph{exposure}), and the wage he/she receives (\emph{outcome}). Specifically, higher intellectual ability increases the probability of graduating from college, and individuals with more ability also tend to earn higher wages. This complication is popularly known as the ``ability bias''~\cite{card1999causal}. The ability bias suggests that standard regression (OLS) coefficient would be a biased estimate of the causal effect of education on earnings.

Over the past decades, researchers have attempted to solve the problem of ``ability bias'' in a number of ways. Notably, a number of studies controlled for the effect of ability bias directly by including measures of ability such as IQ and other test scores within the regression model~\cite{dickson2013causal}. However, there are concerns over whether these types of variables are a good proxy for wage-earning ability. An alternative strategy which has been the focus of much of the literature is to identify one or more variables which affect education but do not affect earnings either directly or indirectly through some other aspect. If such variables can be found, they can be used as \emph{instrumental variables} to derive a consistent estimate of the return to education. A large body of literature has been devoted to identifying proper instruments for estimating the causal effect of education on earnings. Some notable instruments include---differences in education owing to the---proximity to college, quarter-of-birth, and state variation when children have to commence compulsory schooling. A consistent finding across IV studies is that the estimated return to education is 20-40\% above the corresponding OLS estimate~\cite{card1999causal}. These IV studies motivate the question: \emph{could we use IV for quantifying voter bias in Stack Exchange?}

\begin{figure}[hbt]
\centering
\includegraphics[scale=0.35]{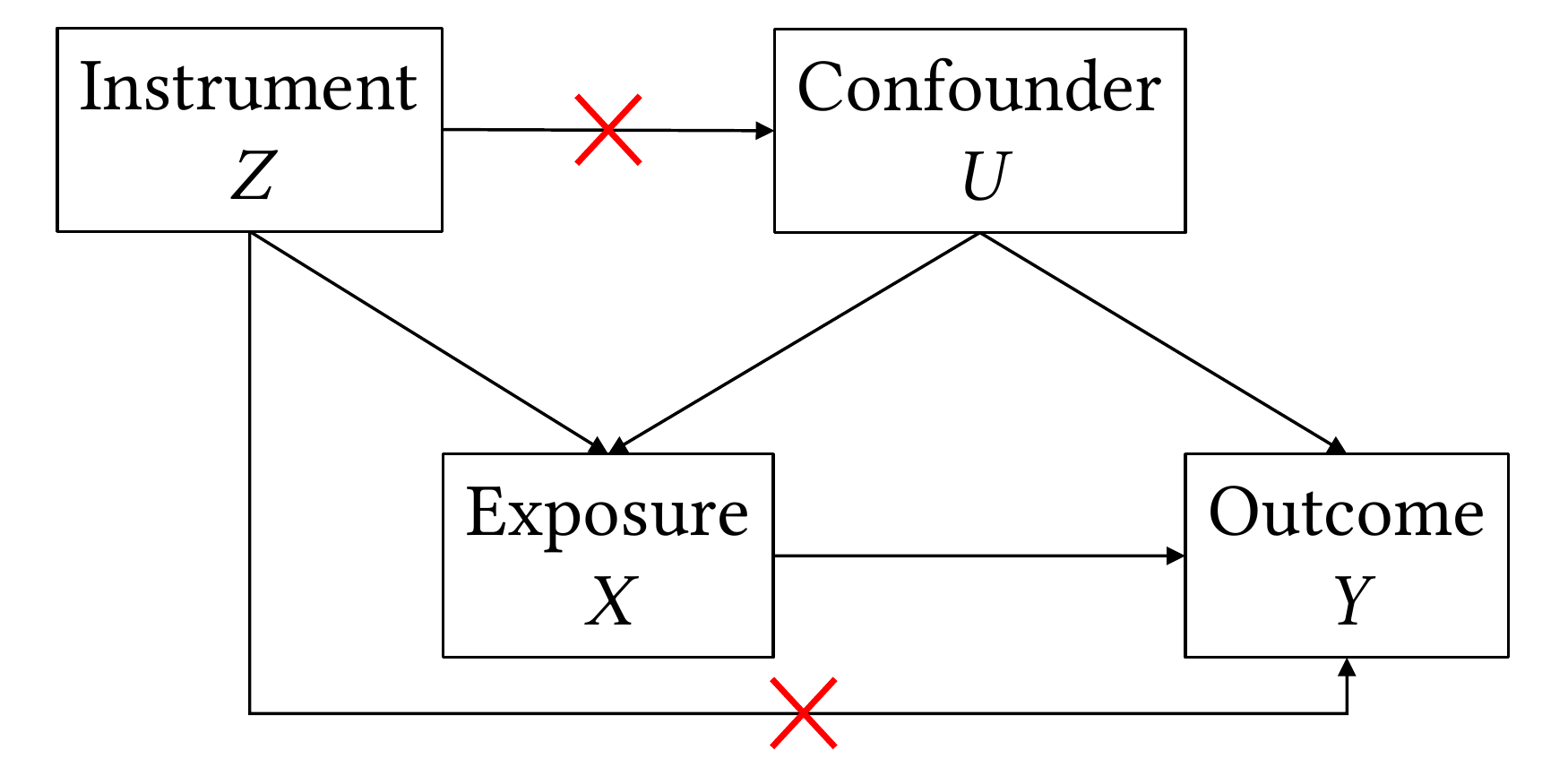}
\caption{General structure of an instrumental variable model. The paths from $U$ to $X$, and $U$ to $Y$ introduces confounding in estimating the causal effect of $X$ on $Y$. For a valid instrument $Z$, the pathways from $Z$ to $X$, and $X$ to $Y$ must exist; whereas the pathways from $Z$ to $U$, and $Z$ to $Y$ must cease to exist.}
\label{fig:iv_structure}
\vspace{-0.25cm}
\end{figure}

Figure~\ref{fig:iv_structure} depicts the general structure of an IV model. Designing an IV model requires identifying a valid \emph{instrument} $Z$---a variable to eliminate the effects of confounders---that must satisfy the following conditions~\cite{hernan2019causal}:

\begin{enumerate}
    \item \emph{Relevance Condition:} The instrument $Z$ is correlated with the exposure $X$. For example, while estimating the causal effect of education on earnings, proximity to college ($Z$) is correlated with college education ($X$).
    \item \emph{Exclusion Restriction:} The instrument $Z$ does not affect the outcome $Y$ directly, except through its potential effect on the exposure $X$. This independence can be conditional upon other covariates. For example, proximity to college ($Z$) should not affect earnings ($Y$), except through its effect on college education ($X$). One can argue that---for people who work at college but are not college graduate themselves---the independence of proximity to college from earnings depends on the job.
    \item \emph{Marginal Exchangeability:} The instrument $Z$ and the outcome $Y$ do not share causes. For example, no common factor influences both proximity to college ($Z$) and earnings ($Y$).
\end{enumerate}

Of the three instrumental conditions mentioned above, only the relevance condition is empirically verifiable~\cite{hernan2019causal}. Therefore, in an observational study such as ours, we can not test if a proposed instrument is a valid instrument. The best we can do is to use our subject matter knowledge to build a case for why a proposed instrument may be reasonably assumed to meet the exclusion restriction and marginal exchangeability.

In IV literature, if the correlation between the instrument $Z$ and the exposure $X$ is strong, then $Z$ is called a \emph{strong instrument}; otherwise, it is called a \emph{weak instrument}. A weak instrument has three major limitations. First, a weak instrument yields parameter estimates with a wide confidence interval. Second, any inconsistency from a small violation of the exclusion restriction gets magnified by the weak instrument. Third, a weak instrument may introduce bias in the estimation process and provide misleading inferences about parameter estimates and standard errors. In this paper, we seek a strong instrument for quantifying each of the three voter biases.

In the remaining subsections, we develop IV models for reputation bias, social influence bias, and position bias. For each voter bias, we operationalize the IV components (outcome, exposure, instrument, and control) using our compiled variables (Table 1).


\subsection{IV Model for Reputation Bias}
\label{ssec:method_3}
We develop an IV model for quantifying reputation bias in Stack Exchange sites. We estimate the causal effect of the reputation of the user who contributes an answer (\emph{exposure}) on the aggregate of votes on that answer (\emph{outcome}). To this end, we operationalize the four IV components (outcome, exposure, instrument, and control) as follows.

\textbf{Outcome.} Our outcome of interest is the aggregate vote on the answer. We represent this outcome via variable \texttt{AnswerScore} \textlangle{}V\textsubscript{19}\textrangle{} \footnote{We shall use this syntax consistently throughout this paper. The first term is variable name and the second term is variable id in Table~\ref{tab:var}. Please see Table~\ref{tab:var} for the description of variables.}.

\textbf{Exposure.} Our exposure of interest is the reputation of the answerer. To represent this exposure, we compute several reputation measures for the answerer, based on the reputation and badge system in Stack Exchange. In Stack Exchange sites, the primary means to gain reputation and badges is to post good questions and useful answers. We compute the reputation measures for each answerer, \emph{per answer}, based on the answerer's achievements prior to creating the current answer. Our reputation measures are as follows: \texttt{AnswererReputation} \textlangle{}V\textsubscript{31}\textrangle{}, \texttt{AnswererReputationViaAnswer} \textlangle{}V\textsubscript{32}\textrangle{}, \texttt{AnswererGoldCount} \textlangle{}V\textsubscript{33}\textrangle{}, \texttt{AnswererSilverCount} \textlangle{}V\textsubscript{34}\textrangle{}, and \texttt{AnswererBronzeCount} \textlangle{}V\textsubscript{35}\textrangle{}.

Note that, for a given answer, different voters may observe different reputation score and badges for the answerer, depending on their time of voting. The voters who participate later typically observe higher reputation score and badges, as the answerer may acquire more upvotes on other answers. Our dataset does not provide the exact state of reputation score and badges of the answerer for a particular vote. To get around this problem, we assume that all voters observe the same state of reputation: the reputation score and badges acquired by the answerer before creating the current answer. In general, reputation increases monotonically; therefore, our assumption is conservative.

Notice that, both our outcome (aggregate votes on the answer) and exposure (reputation of the answerer) of interest can be influenced by the \emph{unobserved ability of the answerer}. Specifically, an answerer with high-ability is expected to generate high-quality answers that would receive many upvotes, increasing his/her reputation. The unobserved ability of the answerer and associated unobserved quality of answers prevent us from distilling the effect of the answerer's reputation on observed votes. We need instruments to eliminate the confounding effect of the answerer's ability.

\textbf{Instrument.} Now, how can we find instruments to uncover the effect of an impression signal (exposure) on the aggregate vote (outcome)? In the social science literature that employs IV's~\cite{gerber1998estimating, jong2005comparative}, researchers use domain knowledge to identify variables that are likely to influence the exposure and thus satisfy the \emph{relevance condition} (these are candidate instruments). Then for each candidate instrument, they use argumentation to determine if it meets the remaining IV conditions---exclusion restriction and marginal exchangeability.

Motivated by the social science approach to IV, we seek candidate instruments that contribute to an answerer's reputation. Based on our literature review, we identify two such factors: 1) answerer's activity level (number of posts, especially answers contributed by the answerer)~\cite{movshovitz2013analysis}, and 2) popularity of the answered questions (number of views, comments, and answers attracted by the questions)~\cite{anderson2012discovering}. Note that an answerer's reputation increases with the volume of his/her activities. Also, a popular question allows contributing answerers to obtain more reputation by attracting more views (voters). To capture these two factors, we compute several measures for each answerer, \emph{per answer}---namely, \texttt{AnswererPostCount} \textlangle{}V\textsubscript{28}\textrangle{}, \texttt{AnswererAnswerCount} \textlangle{}V\textsubscript{29}\textrangle{}, \texttt{AnswererActiveAge} \textlangle{}V\textsubscript{30}\textrangle{}, \texttt{AnsweredQuestionViewTotal} \textlangle{}V\textsubscript{37}\textrangle{}, \texttt{AnsweredQuestionFavoriteTotal} \textlangle{}V\textsubscript{38}\textrangle{}, \texttt{Answered\\QuestionScoreTotal} \textlangle{}V\textsubscript{39}\textrangle{}, \texttt{AnsweredQuestionCommentTotal} \textlangle{}V\textsubscript{40}\textrangle{}, and \texttt{AnsweredQuestion\\AnswerTotal} \textlangle{}V\textsubscript{41}\textrangle{}. We use these variables as are our candidate instruments.

We now scrutinize the candidate instruments to reason about their ability to meet the three instrumental conditions described in Section~\ref{ssec:method_2}. Note that, all three conditions must be met for a candidate instrument to be valid. We divide the candidate instruments into two groups for qualitative reasoning: A) answerer's activity level [\texttt{AnswererPostCount} \textlangle{}V\textsubscript{28}\textrangle{}, \texttt{AnswererAnswerCount} \textlangle{}V\textsubscript{29}\textrangle{}, \texttt{AnswererActiveAge} \textlangle{}V\textsubscript{30}\textrangle{}]; and B) popularity of past questions responded to by the answerer [\texttt{AnsweredQuestionViewTotal} \textlangle{}V\textsubscript{37}\textrangle{}, \texttt{AnsweredQuestionFavoriteTotal} \textlangle{}V\textsubscript{38}\textrangle{}, \texttt{AnsweredQuestion\\ScoreTotal} \textlangle{}V\textsubscript{39}\textrangle{}, \texttt{AnsweredQuestionCommentTotal} \textlangle{}V\textsubscript{40}\textrangle{}, \texttt{AnsweredQuestionAnswerTotal} \textlangle{}V\textsubscript{41}\textrangle{}]. Both groups of candidate instruments empirically satisfy the relevance condition. Therefore, we concentrate on the remaining two IV conditions: exclusion restriction, and marginal exchangeability. In other words, we aim to identify instruments that affect the obtained reputation of the answerer (\emph{exposure}) without affecting the votes on current answer (\emph{outcome}), either directly or through the ability of the answerer (\emph{confounder}).

Notice that the first group of candidate instruments---based on the answerer's activity level---may contribute to the ability of the answerer (\emph{confounder}), which in turn may affect the quality of the answer and resultant votes on the answer (\emph{outcome}). For example, a user who posted many answers may learn from experience to provide better quality answers in the future. Thus, the first group of candidate instruments may violate marginal exchangeability. In contrast, the second group of candidate instruments---based on the popularity of past questions responded to by the answerer---may affect the votes on the current answer (\emph{outcome}) only through the answerer's reputation (\emph{exposure}). These candidate instruments do not inform us about the ability of answerer (\emph{confounder}). The second group of candidate instruments satisfies both exclusion restriction and marginal exchangeability. Ergo, we use the second group of instruments to estimate the effects of reputation signals on observed votes.

\begin{figure}[htb]
\centering
\includegraphics[scale=0.35]{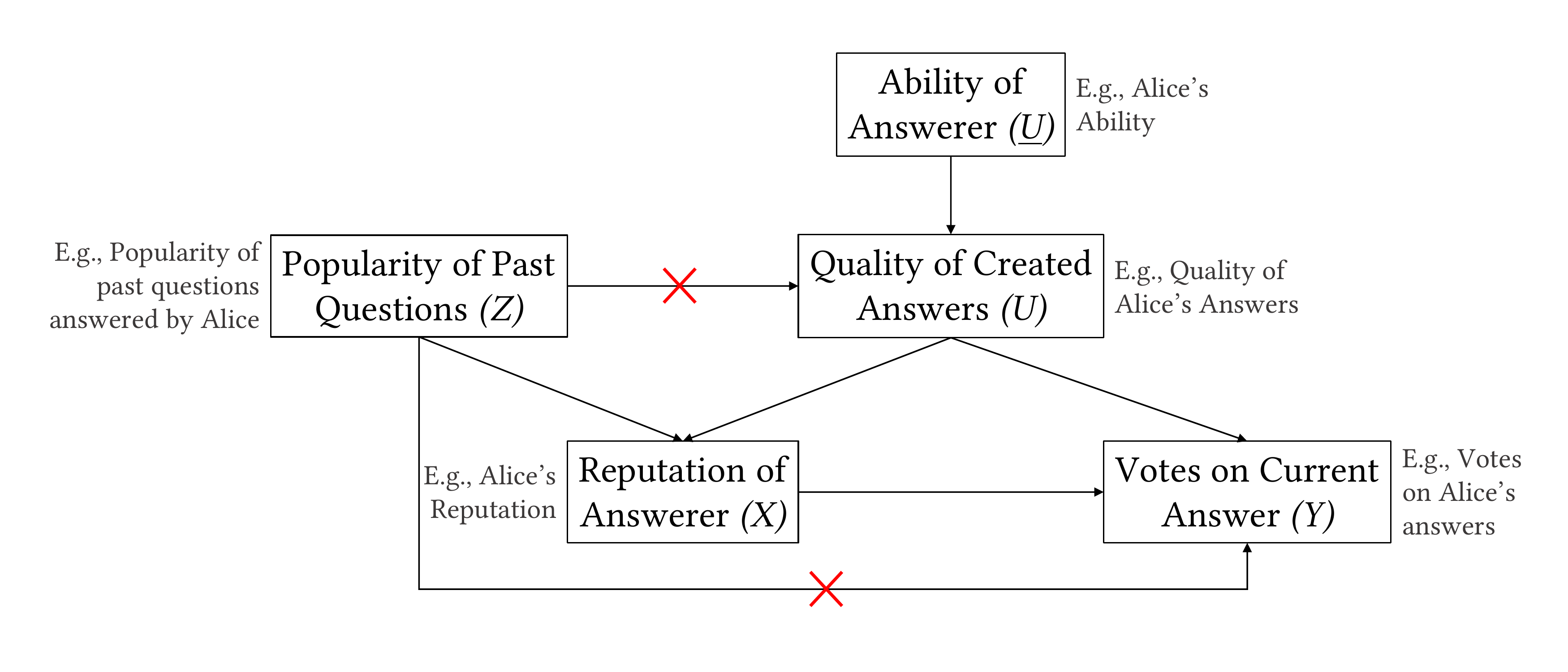}
\caption{Causal diagram of our IV model for quantifying reputation bias. Here, the unobserved ability of answerer introduces confounding via the unobserved quality of created answers. To eliminate this confounding, we propose the popularity of the past questions responded to by the answerer as the instrument.}
\label{fig:reputation_iv}
\vspace{-0.25cm}
\end{figure}

Based on the IV components mentioned above---exposure (reputation of the answerer), outcome (votes on the answer), confounder (the ability of the answerer to create high-quality answers), and instrument (popularity of the past questions)---we present the causal diagram of our model in Figure~\ref{fig:reputation_iv}. Please note that our causal diagram follows the general structure of the instrumental variable framework (in Figure 2).

\textbf{Control.} While our claimed instruments (based on the popularity of past questions responded to by the answerer) are unlikely to affect the outcome (votes on current answer), we take further precautions in the form of controls, to establish the conditional independence of proposed instruments from the outcome. To this end, we propose the following controls in our IV specifiction: \texttt{Site} \textlangle{}V\textsubscript{1}\textrangle{}, \texttt{QuestionViewCount} \textlangle{}V\textsubscript{3}\textrangle{}, \texttt{QuestionFavoriteCount} \textlangle{}V\textsubscript{4}\textrangle{}, \texttt{QuestionScore} \textlangle{}V\textsubscript{5}\textrangle{}, \texttt{QuestionCommentCount} \textlangle{}V\textsubscript{8}\textrangle{}, and \texttt{QuestionAnswerCount} \textlangle{}V\textsubscript{11}\textrangle{}.

Each Stack Exchange site accommodates a distinct audience, who may exhibit a distinct voting behavior; ergo, we control for \texttt{Site} \textlangle{}V\textsubscript{1}\textrangle{} via stratification. The remaining controls capture the popularity of current question, which establish the conditional independence of proposed instruments from the outcome. Specifically, given the popularity of current question, the popularity of past questions responded to by the answerer should not affect the votes on current answer. We incorporate these control variables into our model as regressors. For the outcome (\texttt{AnswerScore} \textlangle{}V\textsubscript{19}\textrangle{}) and exposure of interest (e.g., \texttt{AnswererReputationViaAnswer} \textlangle{}V\textsubscript{32}\textrangle{}), we can select one or more instrumental variables (say \texttt{AnsweredQuestionViewTotal} \textlangle{}V\textsubscript{37}\textrangle{}), and appropriate controls (\texttt{Site} and \texttt{QuestionViewCount} \textlangle{}V\textsubscript{3}\textrangle{}) to estimate the causal effect of the exposure on the outcome.

\subsection{Joint IV Model for Social Influence Bias and Position Bias}
\label{ssec:method_4}
In Stack Exchange sites, the default presentation order of answers is the aggregate vote thus far. This ordering scheme imposes a critical challenge in isolating the effect of position bias from the social influence bias, as the two biases vary together. To address this challenge, we develop a joint IV model to quantify social influence bias and position bias in the same model. We estimate the causal effects of initial votes and resultant position on subsequent votes by specifying the IV components as follows.

\textbf{Outcome.} Our outcome of interest is the aggregate vote on the answer after an initial \emph{bias formation period}---the time required for social influence signal (initial votes) and position signal (answer position) to come into effect. We represent this outcome via \texttt{AnswerScoreT+} \textlangle{}V\textsubscript{21}\textrangle{}: a response variable that captures the aggregate vote on the answer based on the votes after time $T$, where $T$ is the limiting time of bias formation specific to the question.

\textbf{Exposure.} We have two exposures of interest corresponding to the initial votes and resultant position of the answer. To represent these exposures, we compute the aggregate vote and resultant position of answer at the limiting time of bias formation $T$. Our exposures are as follows: \begin{enumerate*} \item \texttt{AnswerScoreT-} \textlangle{}V\textsubscript{20}\textrangle{} captures the aggregate vote on answer based on the votes before time $T$; \item \texttt{AnswerPositionT-} \textlangle{}V\textsubscript{23}\textrangle{} captures the position of answer based on the aggregate vote before time $T$. \end{enumerate*}

\begin{figure}[htb]
\centering
\includegraphics[scale=0.55]{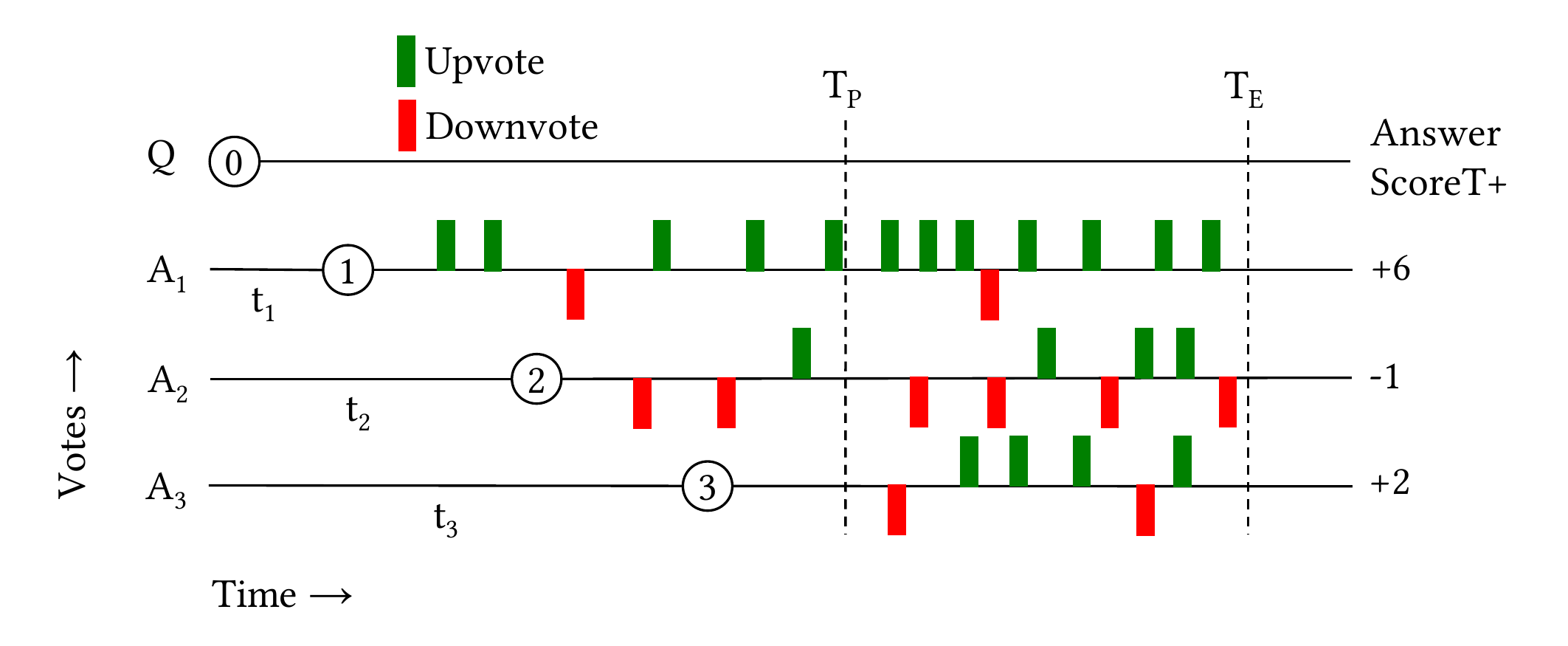}
\caption{An illustration of the bias formation period to quantify our outcome (\texttt{AnswerScoreT+} \textlangle{}V\textsubscript{21}\textrangle{}) and exposures (\texttt{AnswerScoreT-} \textlangle{}V\textsubscript{20}\textrangle{} and \texttt{AnswerPositionT-} \textlangle{}V\textsubscript{23}\textrangle{}). The creation of question $Q$ marks the beginning of our observation period. Then, three answers $A_1$, $A_2$, and $A_3$ that refer to $Q$ arrive after time $t_1$, $t_2$, and $t_3$ respectively. Finally, $T_E$ marks the end of our observation period (the time of data collection). Notice that, a total of 30 votes (20 upvotes, 10 downvotes) are casted on $A_1$, $A_2$, and $A_3$ by time $T_E$. We consider the time by which $P\%$ of total votes are casted on on $A_1$, $A_2$, and $A_3$ as the bias formation period; $T_{P}$ marks the limiting time of this bias formation period. In this example, the value of $P$ is $30$.}
\label{fig:initial_vs_final_votes}
\vspace{-0.25cm}
\end{figure}

We define a bias formation period to quantify our outcome and exposures. We define this period based on the dynamics of votes on the answers to each question. Specifically, we define the bias formation period of a question as the time by which $P\%$ of total votes on its answers are cast. Figure~\ref{fig:initial_vs_final_votes} shows an illustration of bias formation period, and how we use this period to quantify our outcome and exposures. The creation of question $Q$ marks the beginning of our observation period. Then, three answers $A_1$, $A_2$, and $A_3$ that refer to $Q$ arrive after time $t_1$, $t_2$, and $t_3$ respectively. Finally, $T_E$ marks the end of our observation period (the time of data collection). Notice that, a total of 30 votes (20 upvotes, 10 downvotes) are casted on $A_1$, $A_2$, and $A_3$ by time $T_E$. We consider the time by which $P\%$ of total votes are cast on $A_1$, $A_2$, and $A_3$ as the bias formation period; $T_{P}$ marks the limiting time of this bias formation period. In this example, the value of $P$ is $30$ (in our experiments, we use different values of $P$ ranging from 5 to 30). The aggregate vote on answer before time $T_{P}$ is quantified as \texttt{AnswerScoreT-} \textlangle{}V\textsubscript{20}\textrangle{}, and the resultant position as \texttt{AnswerPositionT-} \textlangle{}V\textsubscript{23}\textrangle{}. The values of \texttt{AnswerScoreT-} \textlangle{}V\textsubscript{20}\textrangle{} for answer $A_1$, $A_2$, $A_3$ in Figure~\ref{fig:initial_vs_final_votes} are +4, -1, 0 respectively. The resultant values of \texttt{AnswerPositionT-} \textlangle{}V\textsubscript{23}\textrangle{} for $A_1$, $A_2$, $A_3$ are 1, 3, 2 respectively. The aggregate vote on answer from Time $T_{P}$ to time $T_E$ is quantified as \texttt{AnswerScoreT+} \textlangle{}V\textsubscript{21}\textrangle{}. The values of \texttt{AnswerScoreT+} \textlangle{}V\textsubscript{21}\textrangle{} for $A_1$, $A_2$, $A_3$ are +6, -1, +2 respectively.

Notice that, both our exposures and outcome of interest can be influenced by the \emph{unobserved quality of the answer}. We seek instruments to eliminate the confounding effect of answer quality.

\textbf{Instrument.} We seek candidate instruments that can uncover the effects of initial votes and position on subsequent votes. Same as before, we identify factors that contribute to the initial votes and position, thereby likely to satisfy the \emph{relevance condition}. For the time being, we do not focus on the remaining IV conditions, exclusion restriction and marginal exchangeability. Prior work on voting behavior in Stack Exchange suggest several factors that contribute to initial votes, notably, activities on the question (number of views, comments, and answers attracted by the question)~\cite{anderson2012discovering}, time of answer (day of the week, hour of the day)~\cite{burghardt2017myopia}, and timeliness of answer (time gap between question and answer)~\cite{stoddard2015popularity}. To capture these factors, we compute several measures---namely, \texttt{QuestionScoreT-} \textlangle{}V\textsubscript{6}\textrangle{}, \texttt{QuestionCommentCountT-} \textlangle{}V\textsubscript{9}\textrangle{}, \texttt{QuestionAnswerCountT-} \textlangle{}V\textsubscript{12}\textrangle{}, \texttt{AnswerDayOfWeek} \textlangle{}V\textsubscript{14}\textrangle{}, \texttt{AnswerTimeOfDay} \textlangle{}V\textsubscript{15}\textrangle{}, \texttt{AnswerEpoch}, \texttt{AnswerTimeliness} \textlangle{}V\textsubscript{17}\textrangle{}, and \texttt{AnswerOrder} \textlangle{}V\textsubscript{18}\textrangle{}. These variables are our candidate instruments.

We now scrutinize the candidate instruments to reason about their ability to meet the three instrumental conditions described in Section~\ref{ssec:method_2}. Recall that, all three conditions must be met for a candidate instrument to be valid. We divide the candidate instruments into three groups for qualitative reasoning: A) activities on the question within the bias formation period [\texttt{QuestionScoreT-} \textlangle{}V\textsubscript{6}\textrangle{}, \texttt{QuestionCommentCountT-} \textlangle{}V\textsubscript{9}\textrangle{}, \texttt{QuestionAnswerCountT-} \textlangle{}V\textsubscript{12}\textrangle{}]; B) actual time of answer [\texttt{AnswerDayOfWeek} \textlangle{}V\textsubscript{14}\textrangle{}, \texttt{AnswerTimeOfDay} \textlangle{}V\textsubscript{15}\textrangle{}, \texttt{AnswerEpoch} \textlangle{}V\textsubscript{16}\textrangle{}]; and C) relative timeliness of answer [\texttt{AnswerTimeliness} \textlangle{}V\textsubscript{17}\textrangle{}, \texttt{AnswerOrder} \textlangle{}V\textsubscript{18}\textrangle{}]. All three groups of candidate instruments satisfy the relevance condition. The activities on a question within the bias formation period positively influence the votes on its answers within that period. The actual time of answer creation affects the initial votes due to the varying amount of voter activity across time.The timeliness of an answer affects its initial votes due to the amount time available for voting. Therefore, we concentrate on the remaining two IV conditions: exclusion restriction, and marginal exchangeability. In other words, we aim to identify the instruments that affect the initial votes or position (\emph{exposure}) without affecting the subsequent votes (\emph{outcome}), either directly or through the quality of the answer (\emph{confounder}).

Notice that the first group of candidate instruments---based on the activities on the question within the bias formation period---may be influenced by the popularity of question (\emph{confounder}), which in turn may contribute to both initial votes (\emph{exposure}) and subsequent votes on the answer (\emph{outcome}). For example, a popular question may induce a high amount of activity both within and beyond the bias formation period. The popularity of the question may also explain the initial and subsequent votes on the answer. Thus, the first group of candidate instruments may violate marginal exchangeability. In contrast, the second group of candidate instruments---based on the actual time of answer---may directly influence both initial votes (\emph{exposure}) and subsequent votes on the answer (\emph{outcome}). Thus, the second group of candidate instruments may violate exclusion restriction. Finally, the third group of candidate instruments---based on the relative timeliness of answer---affect the subsequent votes primarily through the initial votes and position. For example, if Bob posts the 2nd answer to a particular question, then his initial votes within the bias formation period will be affected by the fact that he is the 2nd answerer. However, the subsequent votes after the bias formation period will not be affected by the same fact. Note that, the timeliness of an answer may be affected by the answerer's expertise. The answerer's expertise may also affect the outcome (subsequent votes on the answer)~\cite{wu2016role}. We address this issue by incorporating the answerer's expertise as a control variable in our IV model. Notice that, the third group of candidate instruments does not inform us about the quality of the answer (\emph{confounder}) and help us avoid the primary confounder. These candidate instruments are reasonably assumed to satisfy both exclusion restriction and marginal exchangeability. Ergo, we use the third group of instruments to estimate the effects of initial votes and position on subsequent votes.

\begin{figure}[htb]
\centering
\includegraphics[scale=0.35]{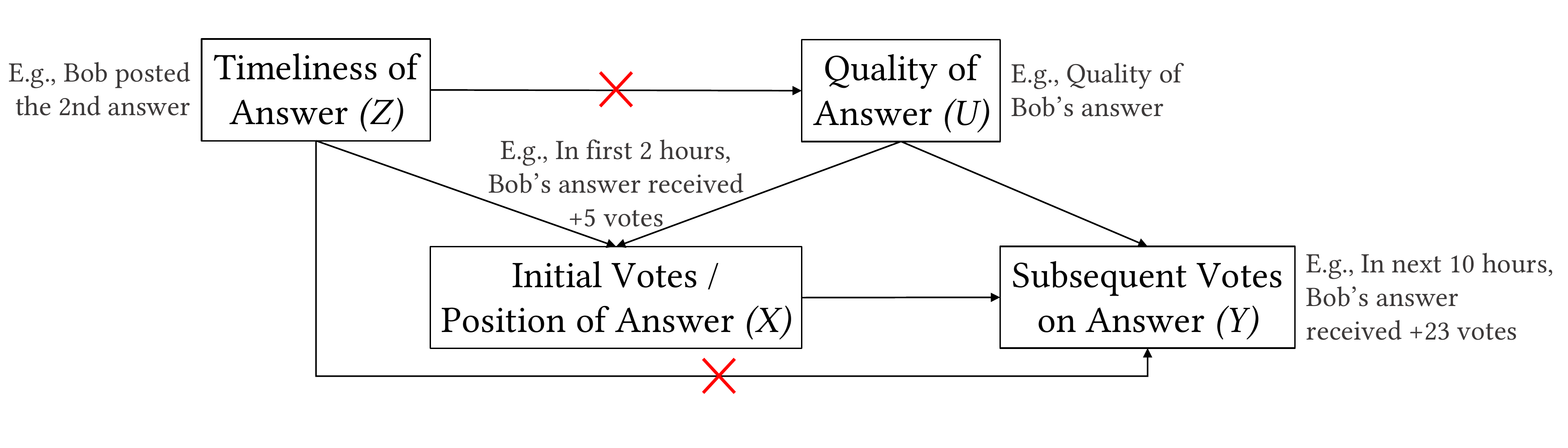}
\caption{Causal diagram of our IV model for quantifying social influence bias and position bias. Here, the unobserved quality of answer act as a confounder. To eliminate this confounder, we propose the timeliness of answer as the instrument.}
\label{fig:social_and_position_iv}
\vspace{-0.25cm}
\end{figure}

Based on the IV components mentioned above---exposure (initial votes and position of the answer), outcome (subsequent votes on the answer), confounder (quality of answer), and instrument (timeliness of answer)---we present the causal diagram of our model in Figure~\ref{fig:social_and_position_iv}.

\textbf{Control.} While our claimed instruments (based on the relative timeliness of answer) are unlikely to affect the outcome (votes on answer after the bias formation period), we take further precautions in the form of controls, to establish the conditional independence of proposed instruments from the outcome. To this end, we propose the following controls in our IV specifiction: \texttt{Site} \textlangle{}V\textsubscript{1}\textrangle{} and \texttt{AnswererReputationViaAnswer} \textlangle{}V\textsubscript{32}\textrangle{}.

In the joint IV model, we control for \texttt{Site} \textlangle{}V\textsubscript{1}\textrangle{} (via stratification) to account for the distinct audience in each Stack Exchange site. We also control for \texttt{AnswererReputationViaAnswer} \textlangle{}V\textsubscript{32}\textrangle{} (via regression) as a proxy for the answerer's expertise. Recall that, our claimed instrument (the timeliness of an answer) may be affected by the answerer's expertise. The answerer's expertise may also affect our outcome (subsequent votes on the answer). While we acknowledge that \texttt{AnswererReputationViaAnswer} \textlangle{}V\textsubscript{32}\textrangle{} is not a proxy for the answerer's expertise, it helps us to reduce the degree of bias in causal estimation.

In this section, we explain how to measure the effects of different impression signals on observed votes through the instrumental variable method. We identify instruments that co-vary only with the impression signals and do not influence the observed through a hidden confounder. These instruments allow us to estimate the causal effect of impression signals on votes.

\section{Results}
\label{sec:results}
In this section, we report the results of our study\footnote{The source code is available at \url{https://github.com/CrowdDynamicsLab/Quantifying_Voter_Biases}}. We begin by presenting the two-stage least squares (2SLS) method for implementing IV models. We then present our bias estimates for Stack Exchange sites---reputation bias (Section~\ref{ssec:results_1}), social influence bias and position bias (Section~\ref{ssec:results_2}).

\textbf{Two-Stage Least Squares (2SLS) Method.} Two-stage least squares (2SLS) is a popular method for computing IV estimates. The 2SLS method consists of two successive stages of linear regression. In the first stage, we regress each exposure variable on all instrumental and control variables in the model and obtain the predicted values from the regressions. In the second stage, we regress the outcome variable on the predicted exposures from the first stage, along with the control variables. The resultant regression coefficients corresponding to the predicted exposures in second stage yield the IV estimates. More details can be found in the supplementary material.

\subsection{Quantifying Reputation Bias}
\label{ssec:results_1}
We quantify reputation bias by estimating the causal effects of reputation score and badges on the aggregate vote. We have one outcome variable (V\textsubscript{19}), five exposure variables (V\textsubscript{31}, V\textsubscript{32}, V\textsubscript{33}, V\textsubscript{34},  V\textsubscript{35}), five instrumental variables (V\textsubscript{37}, V\textsubscript{38}, V\textsubscript{39}, V\textsubscript{40}, V\textsubscript{41}), and six control variables (V\textsubscript{1}, V\textsubscript{3}, V\textsubscript{4}, V\textsubscript{5}, V\textsubscript{8}, V\textsubscript{11}). We use \texttt{Site} \textlangle{}V\textsubscript{1}\textrangle{} to stratify the data based on Stack Exhchange site. We incorporate the remaining variables into 2SLS regression framework to develop our IV models. We develop 10 IV models [5 (exposure) $\times$ 2 (with or without control)] that use all instruments, two for each exposure (with or without control). We develop another 50 IV models [5 (exposure) $\times$ 5 (instrument) $\times$ 2 (with or without control)] to analyze the performance of individual instrument. We also develop a baseline OLS model for each IV model. We perform log modulus transformation [$L(x) = sign(x) * log(|x| + 1\textrangle)$] of variables before using them in regression; this is required to linearize the relationship among variables. The use of log transformation in IV models is well-established~\cite{angrist2010credibility}.

We compare the performance of OLS and IV models by examining their estimates (regression coefficients). Table~\ref{tab:reputation_badges_english},~\ref{tab:reputation_badges_math}, and~\ref{tab:reputation_badges_superuser} present the OLS and IV estimates for quantifying the causal effects of reputation score and badges on the aggregate vote, for \textsc{English}, \textsc{Math}, and \textsc{Superuser} respectively. We make the following observations from these estimates.

\begin{description}

\item [Relevance Condition.] The final instruments for estimating the causal effects of reputation score and badges on the aggregate vote satisfy the \emph{relevance condition} (stated in Section~\ref{ssec:method_2}). For all IV estimates reported in  Table~\ref{tab:reputation_badges_english}--\ref{tab:reputation_badges_superuser}, we observe low $p$-values and high $t$-statistics in the first stage of 2SLS. We do not report these numbers for brevity. Notice that the IV estimates in Table~\ref{tab:reputation_badges_english}--\ref{tab:reputation_badges_superuser} have a small confidence interval, which is a byproduct of identifying \emph{strong instruments}.

\item [Causal Effect of Reputation Score.] Prior research would interpret the regression coefficients from OLS in a causal way. In this paper, we interpret the IV estimates as causal effects. For all three sites, the causal effect of reputation score on the aggregate vote is small. While OLS and IV provide similar estimates for quantifying the effect of reputation score, OLS assigns a slightly higher weight to the reputation score. Control variables rectify the estimates from both OLS and IV by increasing weights. 

\item [Causal Effects of Badges.] For all three sites, the causal effects of badges on the aggregate vote is significant. The effects vary across the level of badges: high effect for gold badges, a moderate effect for silver badges, and low effect for bronze badges. This finding is consistent with the rarity of these badges. Stack Exchange sites grant a few gold badges, some silver badges, and lots of bronze badges to their users. OLS and IV differ a lot in quantifying the effects of badges. OLS tends to assign equal weights to all badges, whereas IV assigns more weight to gold badges (1.6--2.2\texttt{x} of OLS weights). In other words, \emph{OLS underestimates the causal effect of gold badges significantly}. Control variables rectify the estimates from both OLS and IV by increasing weights.

\end{description}

\subsection{Quantifying Social Influence Bias and Position Bias}
\label{ssec:results_2}
We quantify social influence bias and position bias by jointly estimating the causal effects of initial votes and position on the subsequent votes. We have one outcome variable (V\textsubscript{21}), two exposure variables (V\textsubscript{20},  V\textsubscript{23}), two instrumental variables (V\textsubscript{17}, V\textsubscript{18}), and two control variables (V\textsubscript{1}, V\textsubscript{32}). We use \texttt{Site} \textlangle{}V\textsubscript{1}\textrangle{} to stratify the data based on Stack Exchange site. We incorporate the remaining variables into 2SLS regression framework to develop one comprehensive IV model. Note that, we need all instruments and controls to develop our IV model, as there are multiple exposure variables and confounders. For this reason, we can not study the effect of an individual instrument. 


\begin{landscape}
\begin{table}[htb]
\footnotesize\sffamily\mdseries
\caption{Causal effects (regression coefficients) of answerer's reputation score and badges on the aggregate vote in \textsc{English}. All results presented in this table are statistically significant---validated via two-tailed t-tests---with $p < 0.001$. The results suggest that OLS and IV provide similar estimates for reputation score, whereas they differ a lot in estimating the effects of badges. Notably, \emph{OLS tends to assign equal weights to all badges, whereas IV assigns more weights to gold badges}.}
\vspace{-2mm}

\begin{tabular}{@{}llllllll@{}}
\toprule
& Instrument and Control & \multicolumn{5}{c}{\textit{Y} = \textnormal{AnswerScore} \textlangle{}V\textsubscript{19}\textrangle{}} \\ 
\cmidrule{3-7}

& (\emph{for estimating the effect of Exposure}) & \multicolumn{2}{c}{\textit{X} = \textnormal{AnswererReputation} \textlangle{}V\textsubscript{31}\textrangle{}} & & \multicolumn{2}{c}{\textit{X} = \textnormal{AnswererReputationViaAnswer \textlangle{}V\textsubscript{32}\textrangle{}}} \\ 
\cmidrule{3-4} \cmidrule{6-7}

Site & \textit{Z + C} & OLS & IV & & OLS & IV \\ 
\midrule

\textnormal{English} & \textnormal{AnsweredQuestionViewTotal} \textlangle{}V\textsubscript{37}\textrangle{} 
& 0.092 ($\pm$ 0.001) & 0.089 ($\pm$ 0.001) &  & 0.090 ($\pm$ 0.002) & 0.088 ($\pm$ 0.002) \\

& V\textsubscript{37} + \textnormal{QuestionViewCount} \textlangle{}V\textsubscript{3}\textrangle{} 
& 0.101 ($\pm$ 0.002) & 0.098 ($\pm$ 0.001) &  & 0.099 ($\pm$ 0.002) & 0.097 ($\pm$ 0.002) \\

& \textnormal{AnsweredQuestionFavoriteTotal} \textlangle{}V\textsubscript{38}\textrangle{} 
& 0.092 ($\pm$ 0.001) & 0.088 ($\pm$ 0.002) &  & 0.090 ($\pm$ 0.002) & 0.086 ($\pm$ 0.001) \\

& V\textsubscript{38} + \textnormal{QuestionFavoriteCount} \textlangle{}V\textsubscript{4}\textrangle{} 
& 0.101 ($\pm$ 0.002) & 0.093 ($\pm$ 0.001) &  & 0.099 ($\pm$ 0.001) & 0.092 ($\pm$ 0.002) \\

& \textnormal{AnsweredQuestionScoreTotal} \textlangle{}V\textsubscript{39}\textrangle{} 
& 0.092 ($\pm$ 0.001) & 0.086 ($\pm$ 0.002) &  & 0.090 ($\pm$ 0.002) & 0.084 ($\pm$ 0.001) \\

& V\textsubscript{39} + \textnormal{QusestionScore} \textlangle{}V\textsubscript{5}\textrangle{} 
& 0.100 ($\pm$ 0.001) & 0.092 ($\pm$ 0.001) &  & 0.099 ($\pm$ 0.002) & 0.090 ($\pm$ 0.001) \\

& \textnormal{AnsweredQuestionCommentTotal} \textlangle{}V\textsubscript{40}\textrangle{} 
& 0.092 ($\pm$ 0.001) & 0.070 ($\pm$ 0.002) &  & 0.090 ($\pm$ 0.002) & 0.068 ($\pm$ 0.001) \\

& V\textsubscript{40} + \textnormal{QuestionCommentCount} \textlangle{}V\textsubscript{8}\textrangle{} 
& 0.093 ($\pm$ 0.001) & 0.070 ($\pm$ 0.001) &  & 0.091 ($\pm$ 0.002) & 0.069 ($\pm$ 0.002) \\

& \textnormal{AnsweredQuestionAnswerTotal} \textlangle{}V\textsubscript{41}\textrangle{} 
& 0.092 ($\pm$ 0.001) & 0.076 ($\pm$ 0.001) &  & 0.090 ($\pm$ 0.002) & 0.075 ($\pm$ 0.002) \\

& V\textsubscript{41} + \textnormal{QuestionAnswerCount} \textlangle{}V\textsubscript{11}\textrangle{} 
& 0.100 ($\pm$ 0.001) & 0.084 ($\pm$ 0.001) &  & 0.098 ($\pm$ 0.001) & 0.083 ($\pm$ 0.002) \\

& V\textsubscript{37}, V\textsubscript{38}, V\textsubscript{39}, V\textsubscript{40}, V\textsubscript{41} 
& 0.092 ($\pm$ 0.001) & 0.081 ($\pm$ 0.001) &  & 0.090 ($\pm$ 0.002) & 0.079 ($\pm$ 0.002) \\

& V\textsubscript{37}, V\textsubscript{38}, V\textsubscript{39}, V\textsubscript{40}, V\textsubscript{41} + V\textsubscript{3}, V\textsubscript{4}, V\textsubscript{5}, V\textsubscript{8}, V\textsubscript{11}
& 0.098 ($\pm$ 0.002) & 0.087 ($\pm$ 0.001) &  & 0.096 ($\pm$ 0.001) & 0.085 ($\pm$ 0.001) \\
\bottomrule
\end{tabular}

\vspace{3mm}

\begin{tabular}{@{}llllllllll@{}}
\toprule
& Instrument and Control & \multicolumn{8}{c}{\textit{Y} = \textnormal{AnswerScore} \textlangle{}V\textsubscript{19}\textrangle{}} \\
\cmidrule{3-10}

& (\emph{for estimating the effect of Exposure}) & \multicolumn{2}{c}{\textit{X} = \textnormal{AnswererGoldCount} \textlangle{}V\textsubscript{33}\textrangle{}} & & \multicolumn{2}{c}{\textit{X} = \textnormal{AnswererSilverCount} \textlangle{}V\textsubscript{34}\textrangle{}} & & \multicolumn{2}{c}{\textit{X} = \textnormal{AnswererBronzeCount} \textlangle{}V\textsubscript{35}\textrangle{}} \\ 
\cmidrule{3-4} \cmidrule{6-7} \cmidrule{9-10}

Site & \textit{Z + C} & OLS & IV & & OLS & IV & & OLS & IV \\ 
\midrule

\textnormal{English} & \textnormal{AnsweredQuestionViewTotal} \textlangle{}V\textsubscript{37}\textrangle{} 
& 0.184 ($\pm$ 0.006) & 0.712 ($\pm$ 0.014) &  & 0.138 ($\pm$ 0.003) & 0.225 ($\pm$ 0.004) &  & 0.157 ($\pm$ 0.003) & 0.178 ($\pm$ 0.003) \\

& V\textsubscript{37} + \textnormal{QuestionViewCount} \textlangle{}V\textsubscript{3}\textrangle{} 
& 0.219 ($\pm$ 0.005) & 0.794 ($\pm$ 0.014) &  & 0.158 ($\pm$ 0.003) & 0.250 ($\pm$ 0.004) &  & 0.183 ($\pm$ 0.002) & 0.198 ($\pm$ 0.003) \\

& \textnormal{AnsweredQuestionFavoriteTotal} \textlangle{}V\textsubscript{38}\textrangle{} 
& 0.184 ($\pm$ 0.006) & 0.543 ($\pm$ 0.009) &  & 0.138 ($\pm$ 0.003) & 0.187 ($\pm$ 0.003) &  & 0.157 ($\pm$ 0.003) & 0.175 ($\pm$ 0.003) \\

& V\textsubscript{38} + \textnormal{QuestionFavoriteCount} \textlangle{}V\textsubscript{4}\textrangle{} 
& 0.206 ($\pm$ 0.006) & 0.579 ($\pm$ 0.010) &  & 0.153 ($\pm$ 0.002) & 0.200 ($\pm$ 0.003) &  & 0.176 ($\pm$ 0.003) & 0.186 ($\pm$ 0.003) \\

& \textnormal{AnsweredQuestionScoreTotal} \textlangle{}V\textsubscript{39}\textrangle{} 
& 0.184 ($\pm$ 0.006) & 0.570 ($\pm$ 0.010) &  & 0.138 ($\pm$ 0.003) & 0.192 ($\pm$ 0.003) &  & 0.157 ($\pm$ 0.003) & 0.170 ($\pm$ 0.003) \\

& V\textsubscript{39} + \textnormal{QusestionScore} \textlangle{}V\textsubscript{5}\textrangle{} 
& 0.199 ($\pm$ 0.005) & 0.613 ($\pm$ 0.010) &  & 0.151 ($\pm$ 0.003) & 0.207 ($\pm$ 0.003) &  & 0.177 ($\pm$ 0.003) & 0.183 ($\pm$ 0.003) \\

& \textnormal{AnsweredQuestionCommentTotal} \textlangle{}V\textsubscript{40}\textrangle{} 
& 0.184 ($\pm$ 0.006) & 0.447 ($\pm$ 0.010) &  & 0.138 ($\pm$ 0.003) & 0.153 ($\pm$ 0.003) &  & 0.157 ($\pm$ 0.003) & 0.135 ($\pm$ 0.003) \\

& V\textsubscript{40} + \textnormal{QuestionCommentCount} \textlangle{}V\textsubscript{8}\textrangle{} 
& 0.183 ($\pm$ 0.006) & 0.448 ($\pm$ 0.010) &  & 0.138 ($\pm$ 0.003) & 0.154 ($\pm$ 0.004) &  & 0.157 ($\pm$ 0.002) & 0.136 ($\pm$ 0.003) \\

& \textnormal{AnsweredQuestionAnswerTotal} \textlangle{}V\textsubscript{41}\textrangle{} 
& 0.184 ($\pm$ 0.006) & 0.500 ($\pm$ 0.011) &  & 0.138 ($\pm$ 0.003) & 0.170 ($\pm$ 0.003) &  & 0.157 ($\pm$ 0.003) & 0.149 ($\pm$ 0.003) \\

& V\textsubscript{41} + \textnormal{QuestionAnswerCount} \textlangle{}V\textsubscript{11}\textrangle{} 
& 0.201 ($\pm$ 0.006) & 0.551 ($\pm$ 0.010) &  & 0.150 ($\pm$ 0.003) & 0.188 ($\pm$ 0.004) &  & 0.173 ($\pm$ 0.003) & 0.165 ($\pm$ 0.003) \\

& V\textsubscript{37}, V\textsubscript{38}, V\textsubscript{39}, V\textsubscript{40}, V\textsubscript{41} 
& 0.184 ($\pm$ 0.006) & 0.338 ($\pm$ 0.009) &  & 0.138 ($\pm$ 0.003) & 0.143 ($\pm$ 0.003) &  & 0.157 ($\pm$ 0.003) & 0.145 ($\pm$ 0.003) \\

& V\textsubscript{37}, V\textsubscript{38}, V\textsubscript{39}, V\textsubscript{40}, V\textsubscript{41} + V\textsubscript{3}, V\textsubscript{4}, V\textsubscript{5}, V\textsubscript{8}, V\textsubscript{11} 
& 0.195 ($\pm$ 0.005) & 0.382 ($\pm$ 0.008) &  & 0.149 ($\pm$ 0.003) & 0.157 ($\pm$ 0.003) &  & 0.176 ($\pm$ 0.003) & 0.167 ($\pm$ 0.003) \\
\bottomrule
\end{tabular}

\label{tab:reputation_badges_english}
\end{table}
\end{landscape}


\begin{landscape}
\begin{table}[htb]
\footnotesize\sffamily\mdseries
\caption{Causal effects (regression coefficients) of answerer's reputation score and badges on the aggregate vote in \textsc{Math}. All results presented in this table are statistically significant---validated via two-tailed t-tests---with $p < 0.001$. The results suggest that OLS and IV provide similar estimates for reputation score, whereas they differ a lot in estimating the effects of badges. Notably, \emph{OLS tends to assign equal weights to all badges, whereas IV assigns more weights to gold badges}.}
\vspace{-2mm}

\begin{tabular}{@{}llllllll@{}}
\toprule
& Instrument and Control & \multicolumn{5}{c}{\textit{Y} = \textnormal{AnswerScore} \textlangle{}V\textsubscript{19}\textrangle{}} \\ 
\cmidrule{3-7}

& (\emph{for estimating the effect of Exposure}) & \multicolumn{2}{c}{\textit{X} = \textnormal{AnswererReputation} \textlangle{}V\textsubscript{31}\textrangle{}} & & \multicolumn{2}{c}{\textit{X} = \textnormal{AnswererReputationViaAnswer \textlangle{}V\textsubscript{32}\textrangle{}}} \\ 
\cmidrule{3-4} \cmidrule{6-7}

Site & \textit{Z + C} & OLS & IV & & OLS & IV \\ 
\midrule

\textnormal{Math} & \textnormal{AnsweredQuestionViewTotal} \textlangle{}V\textsubscript{37}\textrangle{} 
& 0.056 ($\pm$ 0.001) & 0.055 ($\pm$ 0.001) &  & 0.053 ($\pm$ 0.001) & 0.051 ($\pm$ 0.001) \\

& V\textsubscript{37} + \textnormal{QuestionViewCount} \textlangle{}V\textsubscript{3}\textrangle{} 
& 0.067 ($\pm$ 0.001) & 0.061 ($\pm$ 0.001) &  & 0.063 ($\pm$ 0.001) & 0.057 ($\pm$ 0.001) \\

& \textnormal{AnsweredQuestionFavoriteTotal} \textlangle{}V\textsubscript{38}\textrangle{} 
& 0.056 ($\pm$ 0.001) & 0.057 ($\pm$ 0.001) &  & 0.053 ($\pm$ 0.001) & 0.053 ($\pm$ 0.001) \\

& V\textsubscript{38} + \textnormal{QuestionFavoriteCount} \textlangle{}V\textsubscript{4}\textrangle{} 
& 0.061 ($\pm$ 0.001) & 0.057 ($\pm$ 0.001) &  & 0.058 ($\pm$ 0.001) & 0.053 ($\pm$ 0.001) \\

& \textnormal{AnsweredQuestionScoreTotal} \textlangle{}V\textsubscript{39}\textrangle{} 
& 0.056 ($\pm$ 0.001) & 0.055 ($\pm$ 0.001) &  & 0.053 ($\pm$ 0.001) & 0.051 ($\pm$ 0.001) \\

& V\textsubscript{39} + \textnormal{QusestionScore} \textlangle{}V\textsubscript{5}\textrangle{} 
& 0.058 ($\pm$ 0.001) & 0.053 ($\pm$ 0.001) &  & 0.055 ($\pm$ 0.001) & 0.049 ($\pm$ 0.001) \\

& \textnormal{AnsweredQuestionCommentTotal} \textlangle{}V\textsubscript{40}\textrangle{} 
& 0.056 ($\pm$ 0.001) & 0.040 ($\pm$ 0.001) &  & 0.053 ($\pm$ 0.001) & 0.037 ($\pm$ 0.001) \\

& V\textsubscript{40} + \textnormal{QuestionCommentCount} \textlangle{}V\textsubscript{8}\textrangle{} 
& 0.057 ($\pm$ 0.001) & 0.041 ($\pm$ 0.001) &  & 0.054 ($\pm$ 0.001) & 0.038 ($\pm$ 0.001) \\

& \textnormal{AnsweredQuestionAnswerTotal} \textlangle{}V\textsubscript{41}\textrangle{} 
& 0.056 ($\pm$ 0.001) & 0.040 ($\pm$ 0.001) &  & 0.053 ($\pm$ 0.001) & 0.037 ($\pm$ 0.001) \\

& V\textsubscript{41} + \textnormal{QuestionAnswerCount} \textlangle{}V\textsubscript{11}\textrangle{} 
& 0.060 ($\pm$ 0.001) & 0.043 ($\pm$ 0.001) &  & 0.057 ($\pm$ 0.001) & 0.040 ($\pm$ 0.001) \\

& V\textsubscript{37}, V\textsubscript{38}, V\textsubscript{39}, V\textsubscript{40}, V\textsubscript{41} 
& 0.056 ($\pm$ 0.001) & 0.048 ($\pm$ 0.001) &  & 0.053 ($\pm$ 0.001) & 0.043 ($\pm$ 0.001) \\

& V\textsubscript{37}, V\textsubscript{38}, V\textsubscript{39}, V\textsubscript{40}, V\textsubscript{41} + V\textsubscript{3}, V\textsubscript{4}, V\textsubscript{5}, V\textsubscript{8}, V\textsubscript{11}
& 0.062 ($\pm$ 0.001) & 0.055 ($\pm$ 0.001) &  & 0.059 ($\pm$ 0.001) & 0.050 ($\pm$ 0.001) \\
\bottomrule
\end{tabular}

\vspace{2mm}

\begin{tabular}{@{}llllllllll@{}}
\toprule
& Instrument and Control & \multicolumn{8}{c}{\textit{Y} = \textnormal{AnswerScore} \textlangle{}V\textsubscript{19}\textrangle{}} \\
\cmidrule{3-10}

& (\emph{for estimating the effect of Exposure}) & \multicolumn{2}{c}{\textit{X} = \textnormal{AnswererGoldCount} \textlangle{}V\textsubscript{33}\textrangle{}} & & \multicolumn{2}{c}{\textit{X} = \textnormal{AnswererSilverCount} \textlangle{}V\textsubscript{34}\textrangle{}} & & \multicolumn{2}{c}{\textit{X} = \textnormal{AnswererBronzeCount} \textlangle{}V\textsubscript{35}\textrangle{}} \\ 
\cmidrule{3-4} \cmidrule{6-7} \cmidrule{9-10}

Site & \textit{Z + C} & OLS & IV & & OLS & IV & & OLS & IV \\ 
\midrule

\textnormal{Math} & \textnormal{AnsweredQuestionViewTotal} \textlangle{}V\textsubscript{37}\textrangle{} 
& 0.086 ($\pm$ 0.001) & 0.234 ($\pm$ 0.002) &  & 0.076 ($\pm$ 0.001) & 0.104 ($\pm$ 0.001) &  & 0.090 ($\pm$ 0.001) & 0.112 ($\pm$ 0.001) \\
 
& V\textsubscript{37} + \textnormal{QuestionViewCount} \textlangle{}V\textsubscript{3}\textrangle{} 
& 0.122 ($\pm$ 0.002) & 0.262 ($\pm$ 0.003) &  & 0.094 ($\pm$ 0.001) & 0.116 ($\pm$ 0.001) &  & 0.117 ($\pm$ 0.001) & 0.125 ($\pm$ 0.001) \\

& \textnormal{AnsweredQuestionFavoriteTotal} \textlangle{}V\textsubscript{38}\textrangle{} 
& 0.086 ($\pm$ 0.001) & 0.217 ($\pm$ 0.002) &  & 0.076 ($\pm$ 0.001) & 0.099 ($\pm$ 0.001) &  & 0.090 ($\pm$ 0.001) & 0.115 ($\pm$ 0.001) \\

& V\textsubscript{38} + \textnormal{QuestionFavoriteCount} \textlangle{}V\textsubscript{4}\textrangle{} 
& 0.105 ($\pm$ 0.002) & 0.218 ($\pm$ 0.002) &  & 0.083 ($\pm$ 0.001) & 0.099 ($\pm$ 0.001) &  & 0.103 ($\pm$ 0.001) & 0.115 ($\pm$ 0.001) \\

& \textnormal{AnsweredQuestionScoreTotal} \textlangle{}V\textsubscript{39}\textrangle{}
& 0.086 ($\pm$ 0.001) & 0.214 ($\pm$ 0.002) &  & 0.076 ($\pm$ 0.001) & 0.098 ($\pm$ 0.001) &  & 0.090 ($\pm$ 0.001) & 0.112 ($\pm$ 0.001) \\

& V\textsubscript{39} + \textnormal{QusestionScore} \textlangle{}V\textsubscript{5}\textrangle{}
& 0.100 ($\pm$ 0.001) & 0.206 ($\pm$ 0.002) &  & 0.078 ($\pm$ 0.001) & 0.094 ($\pm$ 0.001) &  & 0.098 ($\pm$ 0.001) & 0.107 ($\pm$ 0.001) \\

& \textnormal{AnsweredQuestionCommentTotal} \textlangle{}V\textsubscript{40}\textrangle{}  
& 0.086 ($\pm$ 0.001) & 0.154 ($\pm$ 0.002) &  & 0.076 ($\pm$ 0.001) & 0.072 ($\pm$ 0.001) &  & 0.090 ($\pm$ 0.001) & 0.081 ($\pm$ 0.001) \\

& V\textsubscript{40} + \textnormal{QuestionCommentCount} \textlangle{}V\textsubscript{8}\textrangle{} 
& 0.089 ($\pm$ 0.002) & 0.157 ($\pm$ 0.002) &  & 0.077 ($\pm$ 0.001) & 0.073 ($\pm$ 0.001) &  & 0.092 ($\pm$ 0.001) & 0.083 ($\pm$ 0.001) \\

& \textnormal{AnsweredQuestionAnswerTotal} \textlangle{}V\textsubscript{41}\textrangle{}
& 0.086 ($\pm$ 0.001) & 0.153 ($\pm$ 0.002) &  & 0.076 ($\pm$ 0.001) & 0.072 ($\pm$ 0.001) &  & 0.090 ($\pm$ 0.001) & 0.081 ($\pm$ 0.001) \\

& V\textsubscript{41} + \textnormal{QuestionAnswerCount} \textlangle{}V\textsubscript{11}\textrangle{} 
& 0.094 ($\pm$ 0.001) & 0.165 ($\pm$ 0.002) &  & 0.081 ($\pm$ 0.001) & 0.077 ($\pm$ 0.001) &  & 0.098 ($\pm$ 0.001) & 0.087 ($\pm$ 0.001) \\

& V\textsubscript{37}, V\textsubscript{38}, V\textsubscript{39}, V\textsubscript{40}, V\textsubscript{41} 
& 0.086 ($\pm$ 0.001) & 0.133 ($\pm$ 0.002) &  & 0.076 ($\pm$ 0.001) & 0.079 ($\pm$ 0.001) &  & 0.090 ($\pm$ 0.001) & 0.092 ($\pm$ 0.001) \\

& V\textsubscript{37}, V\textsubscript{38}, V\textsubscript{39}, V\textsubscript{40}, V\textsubscript{41} + V\textsubscript{3}, V\textsubscript{4}, V\textsubscript{5}, V\textsubscript{8}, V\textsubscript{11} 
& 0.113 ($\pm$ 0.002) & 0.179 ($\pm$ 0.002) &  & 0.085 ($\pm$ 0.001) & 0.090 ($\pm$ 0.001) &  & 0.108 ($\pm$ 0.001) & 0.110 ($\pm$ 0.001) \\
\bottomrule
\end{tabular}

\label{tab:reputation_badges_math}
\end{table}
\end{landscape}


\begin{landscape}
\begin{table}[htb]
\footnotesize\sffamily\mdseries
\caption{Causal effects (regression coefficients) of answerer's reputation score and badges on the aggregate vote in \textsc{Superuser}. All results presented in this table are statistically significant---validated via two-tailed t-tests---with $p < 0.001$. The results suggest that OLS and IV provide similar estimates for reputation score, whereas they differ a lot in estimating the effects of badges. Notably, \emph{OLS tends to assign equal weights to all badges, whereas IV assigns more weights to gold badges}.}
\vspace{-2mm}

\begin{tabular}{@{}llllllll@{}}
\toprule
& Instrument and Control & \multicolumn{5}{c}{\textit{Y} = \textnormal{AnswerScore} \textlangle{}V\textsubscript{19}\textrangle{}} \\ 
\cmidrule{3-7}

& (\emph{for estimating the effect of Exposure}) & \multicolumn{2}{c}{\textit{X} = \textnormal{AnswererReputation} \textlangle{}V\textsubscript{31}\textrangle{}} & & \multicolumn{2}{c}{\textit{X} = \textnormal{AnswererReputationViaAnswer \textlangle{}V\textsubscript{32}\textrangle{}}} \\ 
\cmidrule{3-4} \cmidrule{6-7}

Site & \textit{Z + C} & OLS & IV & & OLS & IV \\ 
\midrule

\textnormal{Superuser} & \textnormal{AnsweredQuestionViewTotal} \textlangle{}V\textsubscript{37}\textrangle{} 
& 0.054 ($\pm$ 0.001) & 0.045 ($\pm$ 0.001) &  & 0.052 ($\pm$ 0.001) & 0.043 ($\pm$ 0.001) \\

& V\textsubscript{37} + \textnormal{QuestionViewCount} \textlangle{}V\textsubscript{3}\textrangle{} 
& 0.067 ($\pm$ 0.001) & 0.062 ($\pm$ 0.001) &  & 0.065 ($\pm$ 0.001) & 0.060 ($\pm$ 0.001) \\

& \textnormal{AnsweredQuestionFavoriteTotal} \textlangle{}V\textsubscript{38}\textrangle{} 
& 0.054 ($\pm$ 0.001) & 0.054 ($\pm$ 0.001) &  & 0.052 ($\pm$ 0.001) & 0.052 ($\pm$ 0.001) \\

& V\textsubscript{38} + \textnormal{QuestionFavoriteCount} \textlangle{}V\textsubscript{4}\textrangle{} 
& 0.065 ($\pm$ 0.001) & 0.062 ($\pm$ 0.001) &  & 0.063 ($\pm$ 0.001) & 0.060 ($\pm$ 0.001) \\

& \textnormal{AnsweredQuestionScoreTotal} \textlangle{}V\textsubscript{39}\textrangle{} 
& 0.054 ($\pm$ 0.001) & 0.052 ($\pm$ 0.001) &  & 0.052 ($\pm$ 0.001) & 0.050 ($\pm$ 0.001) \\

& V\textsubscript{39} + \textnormal{QusestionScore} \textlangle{}V\textsubscript{5}\textrangle{} 
& 0.065 ($\pm$ 0.001) & 0.061 ($\pm$ 0.001) &  & 0.064 ($\pm$ 0.001) & 0.059 ($\pm$ 0.001) \\

& \textnormal{AnsweredQuestionCommentTotal} \textlangle{}V\textsubscript{40}\textrangle{} 
& 0.054 ($\pm$ 0.001) & 0.038 ($\pm$ 0.001) &  & 0.052 ($\pm$ 0.001) & 0.036 ($\pm$ 0.001) \\

& V\textsubscript{40} + \textnormal{QuestionCommentCount} \textlangle{}V\textsubscript{8}\textrangle{} 
& 0.054 ($\pm$ 0.001) & 0.038 ($\pm$ 0.001) &  & 0.052 ($\pm$ 0.001) & 0.036 ($\pm$ 0.001) \\

& \textnormal{AnsweredQuestionAnswerTotal} \textlangle{}V\textsubscript{41}\textrangle{} 
& 0.054 ($\pm$ 0.001) & 0.045 ($\pm$ 0.001) &  & 0.052 ($\pm$ 0.001) & 0.044 ($\pm$ 0.001) \\

& V\textsubscript{41} + \textnormal{QuestionAnswerCount} \textlangle{}V\textsubscript{11}\textrangle{} 
& 0.062 ($\pm$ 0.001) & 0.053 ($\pm$ 0.001) &  & 0.060 ($\pm$ 0.001) & 0.052 ($\pm$ 0.001) \\

& V\textsubscript{37}, V\textsubscript{38}, V\textsubscript{39}, V\textsubscript{40}, V\textsubscript{41} 
& 0.054 ($\pm$ 0.001) & 0.048 ($\pm$ 0.001) &  & 0.052 ($\pm$ 0.001) & 0.046 ($\pm$ 0.001) \\

& V\textsubscript{37}, V\textsubscript{38}, V\textsubscript{39}, V\textsubscript{40}, V\textsubscript{41} + V\textsubscript{3}, V\textsubscript{4}, V\textsubscript{5}, V\textsubscript{8}, V\textsubscript{11}
& 0.063 ($\pm$ 0.001) & 0.060 ($\pm$ 0.001) &  & 0.062 ($\pm$ 0.001) & 0.057 ($\pm$ 0.001) \\
\bottomrule
\end{tabular}

\vspace{2mm}

\begin{tabular}{@{}llllllllll@{}}
\toprule
& Instrument and Control & \multicolumn{8}{c}{\textit{Y} = \textnormal{AnswerScore} \textlangle{}V\textsubscript{19}\textrangle{}} \\
\cmidrule{3-10}

& (\emph{for estimating the effect of Exposure}) & \multicolumn{2}{c}{\textit{X} = \textnormal{AnswererGoldCount} \textlangle{}V\textsubscript{33}\textrangle{}} & & \multicolumn{2}{c}{\textit{X} = \textnormal{AnswererSilverCount} \textlangle{}V\textsubscript{34}\textrangle{}} & & \multicolumn{2}{c}{\textit{X} = \textnormal{AnswererBronzeCount} \textlangle{}V\textsubscript{35}\textrangle{}} \\ 
\cmidrule{3-4} \cmidrule{6-7} \cmidrule{9-10}

Site & \textit{Z + C} & OLS & IV & & OLS & IV & & OLS & IV \\ 
\midrule

\textnormal{Superuser} & \textnormal{AnsweredQuestionViewTotal} \textlangle{}V\textsubscript{37}\textrangle{} 
& 0.106 ($\pm$ 0.004) & 0.414 ($\pm$ 0.009) &  & 0.081 ($\pm$ 0.002) & 0.139 ($\pm$ 0.003) &  & 0.082 ($\pm$ 0.002) & 0.097 ($\pm$ 0.002) \\

& V\textsubscript{37} + \textnormal{QuestionViewCount} \textlangle{}V\textsubscript{3}\textrangle{} 
& 0.175 ($\pm$ 0.004) & 0.591 ($\pm$ 0.009) &  & 0.116 ($\pm$ 0.002) & 0.196 ($\pm$ 0.003) &  & 0.123 ($\pm$ 0.001) & 0.137 ($\pm$ 0.002) \\

& \textnormal{AnsweredQuestionFavoriteTotal} \textlangle{}V\textsubscript{38}\textrangle{} 
& 0.106 ($\pm$ 0.004) & 0.399 ($\pm$ 0.007) &  & 0.081 ($\pm$ 0.002) & 0.143 ($\pm$ 0.002) &  & 0.082 ($\pm$ 0.002) & 0.123 ($\pm$ 0.002) \\

& V\textsubscript{38} + \textnormal{QuestionFavoriteCount} \textlangle{}V\textsubscript{4}\textrangle{} 
& 0.147 ($\pm$ 0.004) & 0.459 ($\pm$ 0.006) &  & 0.103 ($\pm$ 0.001) & 0.165 ($\pm$ 0.002) &  & 0.110 ($\pm$ 0.002) & 0.142 ($\pm$ 0.002) \\

& \textnormal{AnsweredQuestionScoreTotal} \textlangle{}V\textsubscript{39}\textrangle{} 
& 0.106 ($\pm$ 0.004) & 0.406 ($\pm$ 0.007) &  & 0.081 ($\pm$ 0.002) & 0.144 ($\pm$ 0.005) &  & 0.082 ($\pm$ 0.002) & 0.117 ($\pm$ 0.002) \\

& V\textsubscript{39} + \textnormal{QusestionScore} \textlangle{}V\textsubscript{5}\textrangle{} 
& 0.162 ($\pm$ 0.003) & 0.481 ($\pm$ 0.006) &  & 0.109 ($\pm$ 0.001) & 0.170 ($\pm$ 0.002) &  & 0.116 ($\pm$ 0.002) & 0.139 ($\pm$ 0.002) \\

& \textnormal{AnsweredQuestionCommentTotal} \textlangle{}V\textsubscript{40}\textrangle{} 
& 0.106 ($\pm$ 0.004) & 0.266 ($\pm$ 0.006) &  & 0.081 ($\pm$ 0.002) & 0.099 ($\pm$ 0.003) &  & 0.082 ($\pm$ 0.002) & 0.082 ($\pm$ 0.002) \\

& V\textsubscript{40} + \textnormal{QuestionCommentCount} \textlangle{}V\textsubscript{8}\textrangle{} 
& 0.106 ($\pm$ 0.004) & 0.266 ($\pm$ 0.007) &  & 0.081 ($\pm$ 0.002) & 0.099 ($\pm$ 0.003) &  & 0.081 ($\pm$ 0.001) & 0.082 ($\pm$ 0.002) \\

& \textnormal{AnsweredQuestionAnswerTotal} \textlangle{}V\textsubscript{41}\textrangle{} 
& 0.106 ($\pm$ 0.004) & 0.349 ($\pm$ 0.007) &  & 0.081 ($\pm$ 0.002) & 0.124 ($\pm$ 0.002) &  & 0.082 ($\pm$ 0.002) & 0.100 ($\pm$ 0.002) \\

& V\textsubscript{41} + \textnormal{QuestionAnswerCount} \textlangle{}V\textsubscript{11}\textrangle{} 
& 0.144 ($\pm$ 0.003) & 0.419 ($\pm$ 0.007) &  & 0.102 ($\pm$ 0.002) & 0.148 ($\pm$ 0.002) &  & 0.105 ($\pm$ 0.002) & 0.120 ($\pm$ 0.002) \\

& V\textsubscript{37}, V\textsubscript{38}, V\textsubscript{39}, V\textsubscript{40}, V\textsubscript{41} 
& 0.106 ($\pm$ 0.004) & 0.244 ($\pm$ 0.006) &  & 0.081 ($\pm$ 0.002) & 0.110 ($\pm$ 0.002) &  & 0.082 ($\pm$ 0.002) & 0.093 ($\pm$ 0.002) \\

& V\textsubscript{37}, V\textsubscript{38}, V\textsubscript{39}, V\textsubscript{40}, V\textsubscript{41} + V\textsubscript{3}, V\textsubscript{4}, V\textsubscript{5}, V\textsubscript{8}, V\textsubscript{11} 
& 0.152 ($\pm$ 0.003) & 0.337 ($\pm$ 0.005) &  & 0.105 ($\pm$ 0.002) & 0.141 ($\pm$ 0.002) &  & 0.113 ($\pm$ 0.002) & 0.131 ($\pm$ 0.002) \\
\bottomrule
\end{tabular}

\label{tab:reputation_badges_superuser}
\end{table}
\end{landscape}

The measurement of variables in this model relies on the specification of the bias formation period, $T$. We define the bias formation period of a question as the time by which $P\%$ of total votes on its answers are cast. We vary the value of $P$ from 5 to 30, with an increment of 5, to create six different instances of this model. We also develop a baseline OLS instance for each IV instance.

\begin{table}[hbt]
\footnotesize\sffamily\mdseries
\caption{The causal effects (IV estimates) of initial votes and position on subsequent votes in \textsc{English}, \textsc{Superuser} and \textsc{Math}. All results presented in this table are statistically significant---validated via two-tailed t-tests---with $p < 0.001$. The results suggest that OLS and IV differ a lot in quantifying the effects of initial votes and position. Notably, \emph{OLS underestimates reputation bias and overestimates social influence bias significantly}.}
\begin{tabular}{@{}lllllll@{}}

\toprule
& & \multicolumn{5}{c}{\textit{Y} = \textnormal{AnswerScoreT+} \textlangle{}V\textsubscript{21}\textrangle{}, \textit{Z\textsubscript{1}} = \textnormal{AnswerTimeliness} \textlangle{}V\textsubscript{17}\textrangle{}, \textit{Z\textsubscript{2}} = \textnormal{AnswerOrder} \textlangle{}V\textsubscript{18}\textrangle{}} \\ 
\cmidrule{3-7}

& & \multicolumn{2}{c}{\textit{X\textsubscript{1}} = \textnormal{AnswerScoreT-} \textlangle{}V\textsubscript{20}\textrangle{}} & & \multicolumn{2}{c}{\textit{X\textsubscript{2}} = \textnormal{AnswerPositionT-} \textlangle{}V\textsubscript{23}\textrangle{}} \\
\cmidrule{3-4} \cmidrule{6-7}

Site & $T$ & OLS & IV & & OLS & IV \\  
\midrule

\textnormal{English} & $T_{0.05}$ & 0.803 ($\pm$ 0.007) & 0.442 ($\pm$ 0.087) &  & 0.215 ($\pm$ 0.014) & 0.401 ($\pm$ 0.037) \\

& $T_{0.10}$ & 0.821 ($\pm$ 0.006) & 0.403 ($\pm$ 0.080) &  & 0.205 ($\pm$ 0.012) & 0.337 ($\pm$ 0.030) \\

& $T_{0.15}$ & 0.819 ($\pm$ 0.005) & 0.385 ($\pm$ 0.073) &  & 0.184 ($\pm$ 0.010) & 0.300 ($\pm$ 0.025) \\

& $T_{0.20}$ & 0.791 ($\pm$ 0.005) & 0.354 ($\pm$ 0.067) &  & 0.161 ($\pm$ 0.009) & 0.270 ($\pm$ 0.022) \\

& $T_{0.25}$ & 0.752 ($\pm$ 0.004) & 0.323 ($\pm$ 0.061) &  & 0.126 ($\pm$ 0.008) & 0.230 ($\pm$ 0.018) \\

& $T_{0.30}$ & 0.699 ($\pm$ 0.004) & 0.289 ($\pm$ 0.057) &  & 0.100 ($\pm$ 0.008) & 0.204 ($\pm$ 0.016) \\
\midrule

\textnormal{Math} & $T_{0.05}$ & 0.802 ($\pm$ 0.003) & 0.359 ($\pm$ 0.037) &  & 0.470 ($\pm$ 0.007) & 0.483 ($\pm$ 0.010) \\

& $T_{0.10}$ & 0.880 ($\pm$ 0.003) & 0.355 ($\pm$ 0.036) &  & 0.446 ($\pm$ 0.005) & 0.445 ($\pm$ 0.009) \\

& $T_{0.15}$ & 0.920 ($\pm$ 0.003) & 0.352 ($\pm$ 0.035) &  & 0.380 ($\pm$ 0.005) & 0.399 ($\pm$ 0.008) \\

& $T_{0.20}$ & 0.921 ($\pm$ 0.003) & 0.342 ($\pm$ 0.034) &  & 0.339 ($\pm$ 0.004) & 0.373 ($\pm$ 0.007) \\

& $T_{0.25}$ & 0.885 ($\pm$ 0.002) & 0.331 ($\pm$ 0.034) &  & 0.284 ($\pm$ 0.004) & 0.343 ($\pm$ 0.007) \\

& $T_{0.30}$ & 0.833 ($\pm$ 0.002) & 0.324 ($\pm$ 0.033) &  & 0.240 ($\pm$ 0.003) & 0.319 ($\pm$ 0.006) \\ 
\midrule

\textnormal{Superuser} & $T_{0.05}$ & 1.814 ($\pm$ 0.010) & 0.800 ($\pm$ 0.122) &  & 0.842 ($\pm$ 0.025) & 1.209 ($\pm$ 0.058) \\

& $T_{0.10}$ & 1.939 ($\pm$ 0.008) & 0.742 ($\pm$ 0.108) &  & 0.784 ($\pm$ 0.021) & 1.018 ($\pm$ 0.045) \\

& $T_{0.15}$ & 1.983 ($\pm$ 0.007) & 0.689 ($\pm$ 0.097) &  & 0.705 ($\pm$ 0.017) & 0.899 ($\pm$ 0.037) \\

& $T_{0.20}$ & 1.888 ($\pm$ 0.005) & 0.633 ($\pm$ 0.087) &  & 0.594 ($\pm$ 0.014) & 0.793 ($\pm$ 0.030) \\

& $T_{0.25}$ & 1.633 ($\pm$ 0.004) & 0.583 ($\pm$ 0.076) &  & 0.463 ($\pm$ 0.012) & 0.712 ($\pm$ 0.025) \\

& $T_{0.30}$ & 1.477 ($\pm$ 0.003) & 0.526 ($\pm$ 0.067) &  & 0.363 ($\pm$ 0.009) & 0.630 ($\pm$ 0.021) \\ 
\bottomrule

\end{tabular}
\label{tab:joint_english_superuser_math}
\end{table}

We compare the performance of OLS and IV models by examining their estimates (regression coefficients). Table~\ref{tab:joint_english_superuser_math} presents the OLS and IV estimates for quantifying the causal effects of initial votes and position on the subsequent votes, for \textsc{English}, \textsc{Math}, and \textsc{Superuser}. We make the following observations from these estimates.

\begin{description}

\item [Relevance Condition.] The final instruments for estimating the causal effects of initial votes and position on the subsequent votes satisfy the \emph{relevance condition}. For all IV estimates reported in  Table~\ref{tab:joint_english_superuser_math}, we observe low $p$-values and high $t$-statistics in the first stage of 2SLS. We do not report these numbers for brevity. Notice that the IV estimates in Table~\ref{tab:joint_english_superuser_math} have a small confidence interval, which is a byproduct of identifying \emph{strong instruments}.

\item [Causal Effect of Initial Votes.] For all three sites, the causal effect of initial votes on subsequent votes is significant. OLS and IV differ a lot in quantifying the effect of initial votes. OLS assigns high weights to initial votes, 1.8--2.3\texttt{x} of IV weights (based on initial 5\% votes). In other words, \emph{OLS overestimates the causal effect of initial votes significantly}. 

\item [Causal Effect of Initial Position.] For all three sites, the causal effect of initial position on subsequent votes is significant. OLS and IV differ a lot in quantifying the effect of initial position. IV assigns high weights to initial position, at times 1.9\texttt{x} of OLS weights (based on initial 5\% votes). In other words, \emph{OLS underestimates the causal effect of initial position significantly}.

\item [Effect of Bias Formation Period.] For all three sites, increasing the bias formation period $T$ leads to a decrease in causal effects for both initial votes and position. This finding implies that \emph{the first few votes significantly skew the subsequent votes}.

\end{description}

In addition to the above-mentioned definition of bias formation period, we also define it based on the day of question creation. Specifically, we use the votes on answers during the day of question creation for computing \texttt{AnswerScoreT-} \textlangle{}V\textsubscript{20}\textrangle{} and \texttt{AnswerPositionT-} \textlangle{}V\textsubscript{23}\textrangle{}. We use the votes on subsequent days for computing \texttt{AnswerScoreT+} \textlangle{}V\textsubscript{21}\textrangle{}. The results are available in the supplementary material.
\section{Discussion}
\label{sec:discussion}
In the presented work, we quantify the degree of voter biases in online platforms. To derive these bias estimates, we make a methodological contribution in the paper: how to measure the effects of different impression signals on observed votes through a novel application of instrumental variables. Our findings have implications for studying online voting behavior, making changes to the platforms' interface, changes to the policy, and broader research within the CSCW community.

\subsection{Implications for Online Voting Behavior}

Our work has provided some of the first \emph{causal insights} into online voting behavior.

\begin{description}
\item[How Community Type Affects Voting.] Our results show that the effects of impression signals on votes widely vary across Stack Exchange sites. For example, the effect of gold badges in \textsc{English} is twice as high as in \textsc{Math}. Again, the effect of content position in \textsc{Superuser} is twice as high as in \textsc{Math}. This finding implies that what impression signals voters pay attention to and what cognitive heuristics they use to transform the signals into up- and down- votes may vary based upon the community type. For instance, \textsc{English}, \textsc{Superuser}, and \textsc{Math} belong to different themes---culture, technology, and science---which cater to different subsets of participants. On the one hand, different themes induce a varying degree of content interpretation, e.g., content interpretation in \textsc{English} is perhaps more subjective compared to content interpretation in \textsc{Math}~\cite{gkotsis2014s}. On the other hand, users who are interested in different themes may be driven by different factors to contribute~\cite{burghardt2017myopia}. Overall, the communities appropriate the platforms in different ways as they deal with different themes and define their own understanding of what is good content or what signals competent users. Our finding, coupled with the above-mentioned corollaries suggest that voter bias may vary as a function of community type. We follow up on the design implications of these insights in Section~\ref{sub:Informing Policy Design}.

\item[On Social Prestige of Badges.] Our results show that different reputation signals have varying effects on votes. While both badges and reputation score are indicative of user reputation, badges exhibit higher influence on votes compared to reputation score. An interpretation of this finding is that badges are perhaps deemed more ``prestigious'' than reputation score by voters. Recent work by Merchant et al.~\cite{merchant2019signals} investigated the role of reputation score and badges in characterizing social qualities. By adopting a regression approach, they found that reputation score and badges positively correlate with popularity and impact. Our finding, in contrast, provides \emph{causal evidence} in favor of the social prestige of badges~\cite{halavais2014badges}, over reputation score. This evidence, coupled with growing concerns about user engagement in online platforms~\cite{dev2018size} suggest that badge systems may put newcomers at a significant disadvantage. Our results also reveal the relationship between the prestige of badges and their exclusivity. Gold badges are the rarest among the three types of badges, and their effect is \emph{two to three times} higher compared to that of silver and bronze badges.

\end{description}

\subsection{Implications for User Interface Design}
Our research reveals how impression signals in user interface affect the votes and lead to biases. These findings have the potential to inform interface design to avoid biases.

\begin{description}
\item[Conceal Impression Signals.] Our results show that impression signals, such as prior votes and badges, heavily influence voting behavior. An interface design implication of this finding is to conceal these signals from voters. Online platforms may adopt different interface design techniques to conceal impression signals from voters. For example, impression signals can be moved from the immediate vicinity of content; these signals may appear in other places, e.g., badges may still appear in the profile pages of the contributing users. Alternatively, impression signals can be concealed from voters till vote casting; a voter may access the signals only after casting his/her vote. The concept of concealing impression signals has been explored in another context: Grosser et al.~\cite{grosser2014} prescribed removing impression metrics (e.g., number of followers, likes, retweets, etc.) from social media feed to prevent users from feeling compulsive, competitive, and anxious. Note that, while concealing impression signals may eliminate the influence of these signals on voters, it is hard to anticipate how voters will react in the absence of such signals. For instance, voters may then rely on other factors, such as the offline reputation of the contributing user, to make voting decisions. Further, the interface changes may also impact the contributing users, who may adopt new strategic behaviors to maintain their online reputation. 

\item[Delay the Votes.] Recall that, to uncover the effects of prior votes and position on subsequent votes; we use the timeliness of answers as the instrument. The main motivation of our chosen instrument is that early-arriving answers get more time to acquire votes. A design implication of this finding is to prevent the early arriving answer(s) from accumulating higher initial votes. Platforms could withhold the provision of voting for a fixed amount of time to achieve this. The withholding period could be decided based on the historical time gap between the arrival time of questions and answers.

\item[Randomize Presentation Order.] Our results show that the position of content also exhibits a strong influence on voters. As the position of content cannot be concealed in a webpage, the design implication is to eliminate position bias via other means. Platforms may randomize the order of answers for each voter and thus prevent any answer from gaining a position advantage (on average). Lerman et al.~\cite{lerman2014leveraging} studied the effects of different ranking policies on votes, including the randomized ordering policy. They found that random policy is best for unbiased estimates of preferences. However, since a small fraction of user-generated content is interesting, users will mainly see uninteresting content under the random policy. 
\end{description}

\subsection{Informing Policy Design}
\label{sub:Informing Policy Design}

Our research could also inform policy design to mitigate biases.

\begin{description}
\item[De-biasing Votes.] What can a platform operator do to mitigate voting biases? A natural remedy is to de-bias the feedback scores \emph{post-hoc}. Our research provides a major step in this regard by providing accurate bias estimates using the IV approach. Apart from such a remedial approach, platform operators could also use a preventive approach, including adopting more evolved aggregation mechanisms to combine individual feedback from voters. Such complex aggregation already occurs on some websites. For example, Amazon no longer displays the voter average for each product but instead uses a proprietary Machine Learning algorithm to compute the aggregate ratings~\cite{eslami2018communicating}. The aggregation policy for votes may account for potential biases, say by weighting the votes based on their arrival time (later votes are more susceptible to herding behavior), history of the voter (differentiating novice voters from the more experienced voters), and content type. While prior work has considered weighted voting---to identify the answer that received most of the votes when most of the answers were already posted~\cite{romano2013towards}---the weighting mechanism for bias mitigation merits further investigation. It's especially important to understand the effects of weighted voting on participation bias, as different weighting mechanisms may attract different subsets of the voter population to participate. For instance, any weighted voting policy where all votes are not equal is likely to dissuade the disadvantaged voters from participation.

\item[Community Dependent Policy Design.] Our research revealed how community type could affect the degree of voter biases. A policy design implication of this finding is to design policies based on the type of community. Instead of using the same vote aggregation and content ranking function for all Stack Exchange sites, platform operators could use variants of the same function for different sites, accounting for the behavior of the underlying voter base. How variation in policy (across sites) may affect the users who participate in multiple communities is an interesting direction for future research. 
\end{description}

\subsection{Impact on CSCW Research}
\label{sub:Impact on CSCW Research}

We show how to estimate the degree to which a factor bias votes through an application of instrumental variables (IV) method. We believe that IV is a valuable tool for use in CSCW research, in particular, for researchers studying biases and online behavior.

\begin{description}
\item[IV for Studying Biases.] The presented research concentrates on quantifying voter biases in the light of impression signals. However, online platforms also accommodate other more serious forms of biases, such as race and gender biases~\cite{vasilescu2012gender, ford2016paradise, JayBlog}. Jay Hanlon---the vice president of community growth at Stack Overflow---acknowledged the presence of race and gender biases in Stack Exchange: ``Too many people experience Stack Overflow as a hostile or elitist place, especially newer coders, women, people of color, and others in marginalized groups.''~\cite{JayBlog}. Vasilescu et al.~\cite{vasilescu2012gender} revealed the gender representation in Drupal, WordPress, and StackOverflow: only 7-10\% of the participants in these communities are women. Through semi-structured interviews and surveys, Ford et al.~\cite{ford2016paradise} identified some of the barriers for female participation in Stack Overflow, such as lack of awareness about site features and self-doubt about qualification. Estimating the causal effects of race and gender on the perceived community feedback could reveal the degree of race and gender biases in online platforms. We believe IV could be a valuable tool in this regard. The argumentation based underpinning of IV is well-suited for studying biases in observational setup; it prompts researchers to reason about the underlying causal process.
\end{description}

\section{Limitations}
\label{sec:limitations}
The observational nature of our study imposes several constraints on our analysis, which requires us to make a number of assumptions. First, we assume that all voters observe the same state of reputation and badges for the answerer. In reality, voters arrive at different times, and the reputation score and badges of the answerer may change between the voter arrivals. Second, we assume that the voters who arrive after the bias formation period observe the same state of initial votes. However, due to the sequential nature of voting, the observed votes may change from one voter to the next. We also assume that the positions of answers do not change after the bias formation period. Third, we ignore the effects of external influence. For example, a voter may be influenced by Google search results or Twitter promotion to upvote an answer. Fourth, while the default presentation order of answers in Stack Exchange is to sort them by votes, we can not track the views that individuals used to make voting decisions. We assume that the default presentation order is the one that influences voter judgment. Finally, we inherit the key limitation of the instrumental variables method, relying on two untestifiable assumptions: exclusion restriction and marginal exchangeability.

\section{Conclusion}
\label{sec:conclusion}
In content-based platforms, an aggregate of votes is commonly used as a proxy for content quality. However, empirical literature suggests that voters are susceptible to different biases. In this paper, we quantify the degree of voter biases in online platforms. We concentrate on three distinct biases: reputation bias, social influence bias, and position bias. The key idea of our approach is to formulate voter bias quantification using the instrumental variable (IV) framework. The IV framework consists of four components: outcome, exposure, instrument, and control. Using large-scale log data from Stack Exchange sites, we operationalize the IV components by employing impression signals as exposure and aggregate feedback as outcome. Then, we estimate the causal effect of exposure on outcome by using a set of carefully chosen instruments and controls. The resultant estimates quantify the voter biases. Our empirical study shows that the bias estimates from our IV approach differ from the bias estimates from the ordinary least squares (OLS) approach. The implications of our work include: redesigning user interface to avoid voter biases; making changes to platforms' policy to mitigate voter biases; detecting other forms of biases in online platforms.

\printbibliography

\end{document}